\documentclass[]{aa}  

\usepackage{graphicx}
\usepackage{txfonts}
\usepackage{subfig}
\usepackage{textgreek}
\usepackage{multirow}
\usepackage{mathtools}
\usepackage{color}
\usepackage{lmodern}
\usepackage{epstopdf}

\bibpunct{(}{)}{;}{a}{}{,}

\newcommand{\FIG}[1]{}

\def\mso{\,{\rm M}_\odot}

\def\lso{\,{\rm L}_\odot}

\def\kms{\, {\rm km}\, {\rm s}^{-1}}

\def\msoy{\, \mso~{\rm yr}^{-1}}

\begin{document}

% ***********************************************************************************************************************************************
%                                                                   TITLE PAGE
% ***********************************************************************************************************************************************

   \title{An unusual face-on spiral in the wind of the M-type AGB star EP Aquarii. }

   \author{Ward Homan
          \inst{1}
          \and
          Anita Richards
          \inst{2}
          \and
          Leen Decin
          \inst{1,3}
          \and
          Alex de Koter
          \inst{1,5}
          \and
          Pierre Kervella
          \inst{4}
          }

   \offprints{W. Homan}          
          
   \institute{$^{\rm 1}\ $Institute of Astronomy, KU Leuven, Celestijnenlaan 200D B2401, 3001 Leuven, Belgium \\
             $^{\rm 2}\ $JBCA, Department Physics and Astronomy, University of Manchester, Manchester M13 9PL, UK \\
             $^{\rm 3}\ $University of Leeds, School of Chemistry, Leeds LS2 9JT, United Kingdom \\
             $^{\rm 4}\ $LESIA (CNRS UMR 8109), Observatoire de Paris, PSL, CNRS, UPMC, Univ. Paris-Diderot, France \\
             $^{\rm 5}\ $Sterrenkundig Instituut `Anton Pannekoek', Science Park 904, 1098 XH Amsterdam, The Netherlands\\
             }

   \date{Received <date> / Accepted <date>}
 
   \abstract  
   {High-resolution interferometric observations of the circumstellar environments of AGB stars show a variety of morphologies. Guided by the unusual carbon monoxide line profile of the AGB star EP Aquarii, we have observed its circumstellar environment with ALMA band 6 in cycle 4. We describe the morphological complexity of the CO, SiO, and SO$_{\rm 2}$ molecular emission. The CO emission exhibits the characteristics of a bi-conical wind with a bright nearly face-on spiral feature around the systemic velocity. This is the first convincing detection of a spiral morphology in an O-rich wind. Based on the offsets  of the centres of the two bi-conical wind hemispheres, we deduce the position angle of the inclination axis to be $\sim$150$^\circ$ measured  anticlockwise from north. Based on the velocity width of the spiral signature, we estimate the inclination angle of the system to be between 4$^\circ$ and 18$^\circ$. The central emission zone exhibits a morphology that resembles simulations modelling the spiral-inducing wind Roche-lobe overflow mechanism. Though the spiral may be a companion-induced density enhancement in the stellar outflow, the extremely narrow width of the spiral signature in velocity space suggests that it may be a hydrodynamical perturbation in a face-on differentially rotating disk. The SiO emission does not show the spiral, but exhibits a local emission void approximately 0.5'' west of the continuum brightness peak. We hypothesise that this may be a local environment caused by the presence of a stellar companion with a mass of at most 0.1$\mso$, based on its non-detection in the continuum. Finally, the SO$_{\rm 2}$ emission remains confined to a 0.5'' radius, and does not show any obvious substructure, but it exhibits a clear rotation signature. Combined, the properties of the molecular emission favour the face-on rotating disk scenario. We observe unexpectedly large red- and blue-shifted  wings in the spectral line of SiO, which could be explained by the potential non-local thermodynamic equilibrium (NLTE) nature of driven, mixed, partly granular fluids.}
   
   \keywords{Radiative transfer--Stars: AGB and post-AGB--circumstellar matter--Submillimeter: stars--Molecular data}

   \maketitle

% ***********************************************************************************************************************************************
%                                                                   END TITLE PAGE
% ***********************************************************************************************************************************************

% ===================================================================================================================================================

% \tableofcontents

\section{Introduction}

Asymptotic giant branch (AGB) stars, named after their position in the Hertzsprung--Russell diagram, are the evolutionary stage of low- and intermediate-mass stars before they turn into the central star of their extended nebular progeny, the planetary nebulae. This metamorphosis originates from the intense mass loss these AGB stars undergo, believed to originate from radiation pressure on dust grains \citep{Winters2000,Hofner2003,Woitke2006}. This dust is formed under the suitable  conditions (high-density, low-temperature) in the nearby circumstellar environment (CSE). The incident stellar radiation couples with the opacity of the dust grains, propelling them outwards, and dragging the circumstellar gas along. The shed material forms extended CSEs which harbour a wide spectrum of complex interlinked thermal, radiative, chemical, and dynamical processes.

These CSEs exhibit large-scale sphericity, but generally display a rich spectrum of smaller scale structural complexities. These include bipolar structures \citep[e.g.][]{Balick2013}, arcs \citep[e.g.][]{Decin2012,Cox2012}, shells \citep[e.g.][]{Mauron2000}, clumps \citep[e.g.][]{Bowers1990,Ohnaka2016,Khouri2016}, spirals \citep[e.g.][]{Mauron2006,Mayer2011,Maercker2012,Kim2013}, tori \citep[e.g.][]{Skinner1998}, voids \citep{Ramstedt2014}, and rotating disks \citep{Kervella2016}. Current hydrodynamic models show that wind--companion interactions, where the wind of the AGB star is perturbed by the local gravity field of a non-giant binary companion, may play a crucial role in explaining various deviations from spherical symmetry \citep[e.g.][]{Soker1997,Huggins2007,Mastrodemos1999,Kim2011,Kim2012}. This does not come as a surprise, as the multiplicity frequency of the main sequence (MS) predecessors of AGB stars is found to exceed 50\%\ \citep{Raghavan2010,Duchene2013}, not including any planetary companions. Considering that recent studies show that on average every star in the Milky Way \citep{Cassan2012} possesses one or more planets, this frequency can only be considered a lower limit.

Thanks to the currently available high spatial resolution facilities, the winds of AGB stars exhibiting any deviation from spherical symmetry can now be studied to much finer detail. Resolving the CSE structure on, for instance, the scale of the binary separation provides valuable insights on the local dynamics dominating the wind--binary interaction zone and dictates the global shape and evolution of the CSE. High-resolution images are thus able to shed light on the formation mechanism of the above-mentioned morphologies. Revealing and understanding the local inner wind dynamics sheds light on the processes that dictate the further evolution into post-AGB stars and planetary nebulae, whose morphologies have been extensively documented and have been found to be mostly highly aspherical. Therefore, an in-depth understanding of the finer physical details governing the inner regions of complexity-harbouring AGB CSEs will aid in the quest to understand the missing morphological link between AGB stars and their progeny.

This study focuses on the AGB star EP Aquarii. EP Aqr is an oxygen-rich, M-type, semi-regular variable AGB star (spectral type M8III) with a period of 55 days \citep{Lebzelter2002}. Hipparcos distance measurements put it at a distance of approximately 135 parsec \citep{Lebzelter2002}. \citet{Dumm1998} estimated an effective temperature of $T_*\simeq$ 3200K and a mass of $M_*\simeq1.7\,\mso$. Its estimated luminosity is $\sim$ 4800$\,\lso$. The object has a radial velocity relative to the local standard of rest (lsr) of $v_{*} = +33.98\,\kms$, a  mass-loss rate of $\sim$1.2$\times$10$^{\rm -7}\,\msoy$, and a terminal wind velocity of $\sim$11$\,\kms$ \citep{Nhung2015}. \citet{Winters2007} have combined interferometric and single-dish data to make a spectral map of the emission from the inner ($10'' \times 10''$) wind, and found that the spectral emission of the CO rotational transitions $J$=1$-$0 and $J$=2$-$1 show a peculiar dual nature, with a broad emission feature defining the width of the line, superposed by a bright, narrow central peak, much like the line profile of L$_{\rm 2}$ Puppis \citep{Winters2002}. The shape of the emission on offset positions shows the source of this peculiarity to be extended (Fig. 3 in Winters et al. 2007). Furthermore, the position--velocity (PV) diagrams constructed from the interferometric data\footnote{PdB Interferometer (Plateau de Bure); the single dish IRAM 30m (Pico Veleta).} reveal a pattern that closely resembles the synthetic PV diagrams generated by inclined flattened spiral structures in smooth spherical outflows \citep{Homan2015} and those of equatorial density enhancements with polar biconical outflows \citep{Homan2016}.

To further refine the analysis by \citet{Winters2007}, in the autumn of 2016 we obtained spatially resolved cycle 4 ALMA band 6 CO, SiO, and SO$_{\rm 2}$ observations of the CSE of EP Aquarii using three different antenna configurations  (see Sect. \ref{obs}). In the present work we present all morphological features identified in the ALMA data. In particular, we focus on the extended CO emission and on the more compact SiO and SO$_{\rm 2}$ emission. The data shows a relatively classical outflow, with a small-scale clumpy and filamentous structure. Around central velocity the CO emission exhibits a strong increase in brightness, and reveals a prominent, finely resolved, almost face-on spiral. This is the first convincing detection of a spiral in an oxygen-rich environment. In addition, the SiO molecular emission exhibits what we believe could be a stellar absorption feature to the west of the AGB continuum peak, potentially revealing the location and nature of the spiral-inducing binary companion.

The paper is organised as follows. In Sect. \ref{obs} we describe the ALMA observation conditions, the data reduction, and the combination of the data observed at different epochs with different antenna configurations. In Sect. \ref{datadescription},  we present all the identified morphological complexities in the CO, SiO, and SO$_{\rm 2}$ molecular emission. Subsequently, in Sect. \ref{discus}, we discuss the potential origins of the observed morphologies and deduce some geometrical properties of the system. We also consider a possible origin for the substantial discrepancy between the spectral line widths of the CO and SiO emission. Finally, in Sect. \ref{summ}, we summarise our findings. For best visual appearance all figures are should be viewed on screen.

\begin{table*}
        \caption{EP Aquarii ALMA observation: Antenna configuration and observation details. The precipitable water vapour (pvw) was $\sim$1.2 mm for each observation.}
        \centering          
        \label{obs1}
        \begin{tabular}{lcllcclc}
        \hline\hline
        \noalign{\smallskip}
Array & Number of & Baselines &  Synthesised & Beam PA & MRS & Date of & Time on \\
& antennas & (m) & beam size ('') & ($^\circ$) & ('') & observation & target (min) \\
        \noalign{\smallskip}
        \hline    
        \noalign{\smallskip}
ACA (7m) & 9 & 9-49 & 9.20$\times$3.80 & -80 & 17 & 2016-10-08 & 43 \\
        \hline    
        \noalign{\smallskip}
TM1 (12m) & 41 & 19-3100 & 0.16$\times$0.14 & -25 & 7& 2016-10-08 to 09 & 170 \\
        \hline    
        \noalign{\smallskip}
TM2 (12m) & 42 & 19-492 & 0.80$\times$0.66 & -77 & 7 & 2016-12-13 to 17 & 60 \\
        \hline
      \end{tabular}
\end{table*}

\begin{table*}
        \caption{Selected ALMA spectral windows in the combined data. The sky frequencies assume a source radial velocity of $v_{\rm lsr} = -33.98$\,km\,s$^{-1}$. The central observed frequency is the average over epochs.}
        \centering          
        \label{spw}
        \begin{tabular}{clllrccr}
        \hline\hline
        \noalign{\smallskip}
spw & Objective & Transition & Observed $\nu$ & Chan. width & Velocity res. & Noise rms & Beam\\
& & & (GHz) & (kHz) & (km\,s$ ^{-1}$) & (mJy) & (arcsec $\times$ arcsec, deg)\\
        \noalign{\smallskip}
        \hline    
        \noalign{\smallskip}
%        Baseband 1:\\
0 & $^{12}$CO & $J$=2$-$1 & 230.55 & 244.141 & 0.317 & 1.3 & $0.18\times0.17$, -15 \\
        \hline    
        \noalign{\smallskip}
%        Baseband 2:\\
1 & NaCl & $J$=18$-$17 & 234.27 & 488.281 & 0.625 & 1.1 & $0.18\times0.17$, -12 \\
        \hline    
        \noalign{\smallskip}
%        Baseband 3:\\
2 & SiO & $J$=5$-$4 & 217.12 & 244.141 & 0.337 & 1.2 & $0.20\times0.18$, +8 \\
        \hline    
        \noalign{\smallskip}
%        Baseband 4:\\
3 & SiS & $J$=12$-$11 & 217.83 & 244.141 & 0.336 & 1.1 & $0.20\times0.18$, -20 \\
        \hline
      \end{tabular}
\end{table*}

\begin{figure*}[]
        \centering
        \includegraphics[width=8cm]{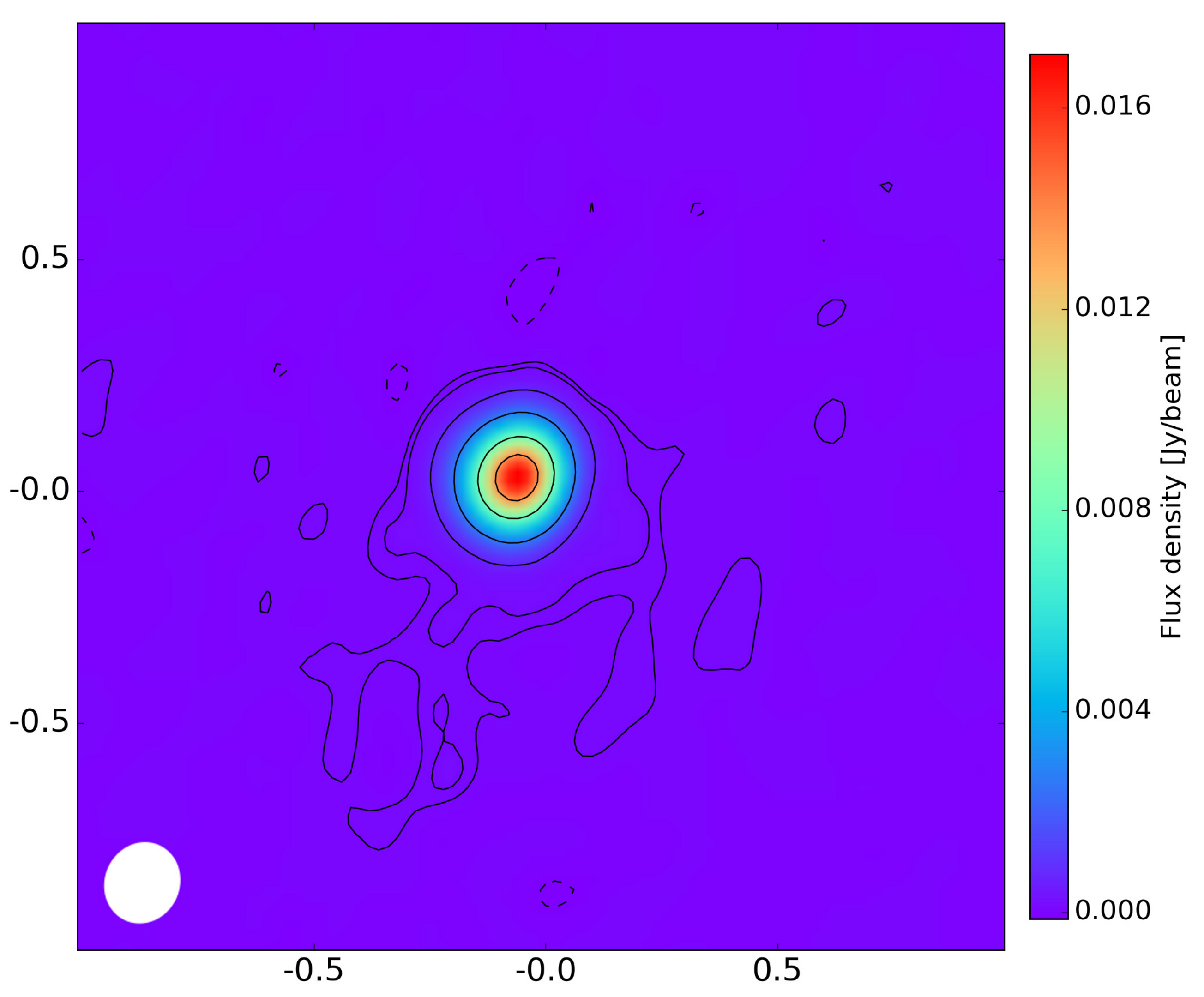}
        \includegraphics[width=9cm]{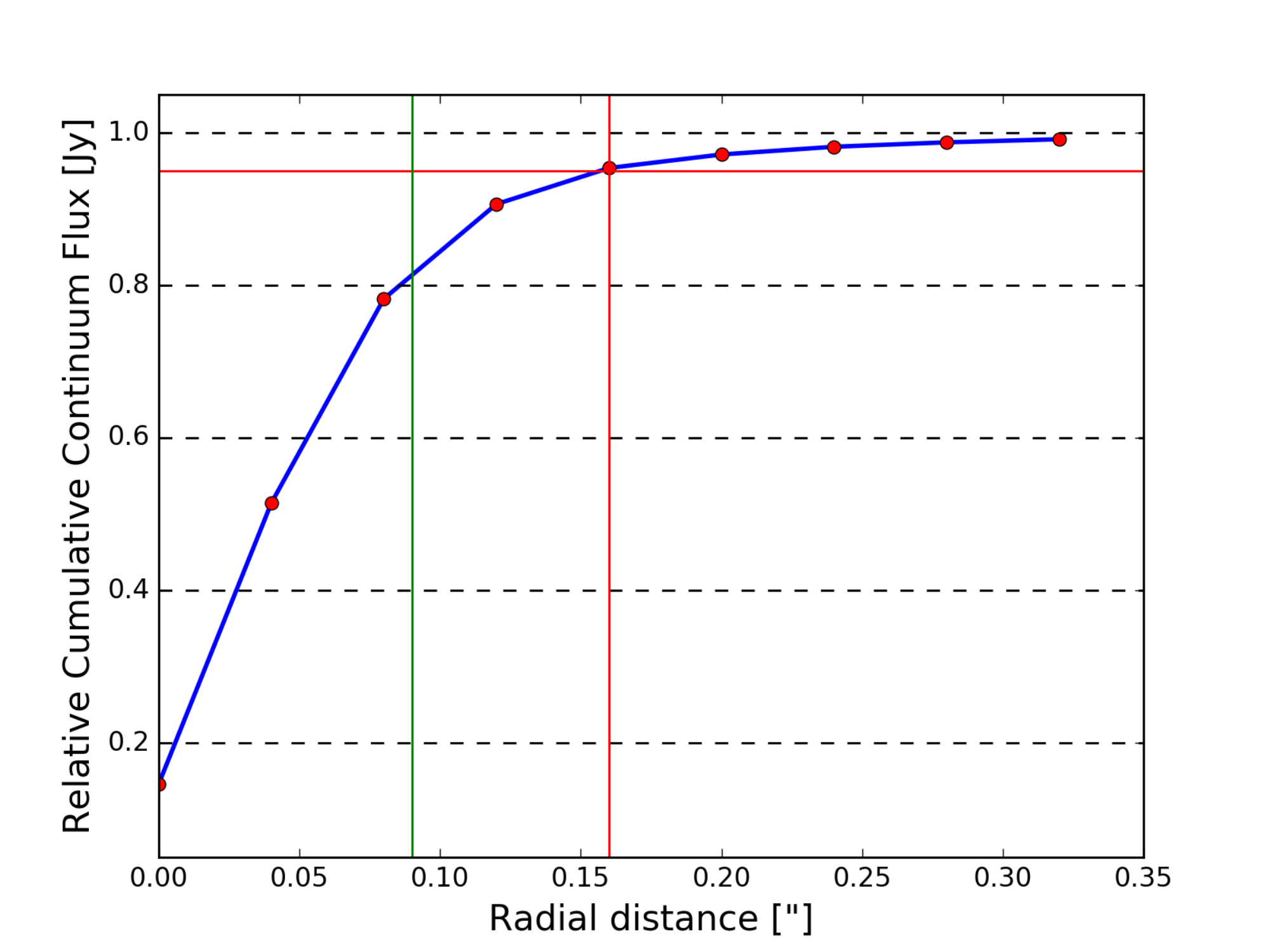}
        \caption{\emph{Left}: Continuum emission of the EP Aqr system. Contours are drawn at -3 (dashed), 3, 5, 25, 100, 300, and 500 times the continuum rms noise value (2.77 $\times {\rm 10}^{\rm -5}$ Jy/beam). The ALMA beam size is shown in the bottom left corner. \emph{Right}: Azimuthally averaged normalised cumulative continuum flux distribution. The reference flux is the total continuum flux, amounting to 1.39 Jy. The red lines indicate the border of the region containing 95\% of the total flux, at 0.16''. The green line indicates the radius of half the FWHM of the resolving beam.
        \label{cont}} 
\end{figure*}

\begin{table*}
        \caption{Molecular emission lines detected in EP Aquarii.}
        \centering          
        \label{idline}
        \begin{tabular}{lcclr}
        \hline\hline
        \noalign{\smallskip}
        Molecule & \multicolumn{2}{c}{Quantum numbers}  & Rest freq. $\nu_0$ & Upper-state \\
         & Vibrational & Rotational  & (GHz) & energy (K) \\
        \noalign{\smallskip}
        \hline    
        \noalign{\smallskip}
$^{12}$CO & $\varv=0$ & $2-1$ & 230.53800 & 16.60 \\
$^{28}$SiO & $\varv=0$ & $5-4$ & 217.10498 & 31.26 \\
SO$_{\rm 2}$ & $\varv=0$ & $\rm{28}_{\rm 3,25}-\rm{28}_{\rm 2,26}$ & 234.42160 & 213.32 \\
        \hline                      
        \end{tabular}
\end{table*}

\section{Data acquisition and reduction} \label{obs}

\subsection{ALMA observations}

EP Aquarii was observed by ALMA in Band 6 for project code 2016.1.00057.S (PI W. Homan). The star was observed with three different antenna configurations (details in Table~\ref{obs1}. The ACA observations used a single pointing at a right ascension of 21:46:31.8750  and a declination $-$02:12:45.604; the 12m observations used a triangularly symmetric three-point mosaic with the combined centre corresponding to the ACA position. The total field of view to the half-power primary beam is approximately $46\arcsec$ for the 7m configuration, and approximately $33\arcsec$ for the 12m mosaics. Because the observations could not be carried out simultaneously, the proper motion of EP Aqr (24.98$\pm$0.75, 19.54$\pm$0.33) mas yr$^{\rm -1}$ \citep{VanLeeuwen2007} was taken into account. The observations used spectral windows (spw) placed as described in Table~\ref{spw}. The effective width of each spw (after shifting to constant $v_{*}$ and selecting overlapping regions) is $\sim$0.46 GHz. Each spw was centred on a spectral line of interest (see Table~\ref{idline}), the frequencies being adjusted for each observing session for the $v_{*}$ of the target at that epoch. Channel averaging of 2 or 4 was applied in the correlator so despite Hanning smoothing, the output channel spacing corresponds to the effective spectral resolution.

Bandpass and time-dependent calibration used compact QSOs. The observed QSO fluxes and their roles are found in Table \ref{qso} in the Appendix. The phase-reference sources used were within 4 degrees of the target. Neptune was used as the flux scale standard for the ACA observations, and the flux densities of the standard QSOs used for the 12m observations are established with respect to Neptune by regular monitoring. The standard flux scales are accurate to about 5\%, and the overall accuracy is about 7\% after using  these values to derive the flux densities of the other sources.

\subsection{Data reduction}

% After applying the phase-reference and other calibrations to L$_2$\,Pup, the corrected target data were split out and each spectral window adjusted to fixed velocity with respect to the LSR. Obvious spectral lines were identified in the visibility data, leaving 2.5\,GHz of line-free continuum.
% A copy of the data with all channels averaged to the coarsest resolution was made to speed up continuum imaging.
% The continuum image (made with natural weighting) has a synthesized beam size $17.7 \times 14.5$\,mas at position angle (PA) 73$^\circ$.
% 
% The position of the continuum peak was located at $\alpha = 07$:13:32.47687, $\delta = -44$:38:17.8443 with an absolute position uncertainty of $\pm 7$\,mas.
% The clean components of this image were used as a model for phase self-calibration, and iterative cycles of phase and amplitude self-calibration were performed.
% Multi-frequency synthesis was used with a linear position-dependent spectral index as a free parameter; although the spectral index is not reliable except for the brightest emission over the relatively narrow, unevenly sampled bandwidth, this improves the image fidelity.
% A 2.5\,mas pixel size and a field of view of $2.56\arcsec$ were used for all images unless otherwise stated.

The ACA and TM2 data were calibrated using the ALMA pipeline; the TM1 data were manually calibrated using standard scripts (see the ALMA Technical Handbook\footnote{https://www.iram.fr/IR.A.MFR/ARC/documents/cycle4/\\ALMA\_Cycle4\_Technical\_Handbook-Final.pdf} for information, or \citealt{Kervella2016}, where similar procedures have been followed). For each configuration, we extracted the target and adjusted the data to constant velocity in the direction of the Local Standard of Rest ($v_{*}$). Copies of the TM1 and TM2 data were made with channel averaging to 15.625 MHz to speed up continuum imaging. We inspected the visibility spectra for each configuration to identify the line-free channels, and these continuum channels were imaged. The phase stability on the longest baselines (TM1) was improved by phase self-calibration of the continuum, with the solutions applied to all data.

All produced images were centred on the nominal position for epoch 2016-10-08. Any offsets for the more compact configurations were less than the uncertainties. The actual positions measured for the peak of EP Aqr at the highest accuracy agreed to within less than 1 mas, at right scension 21:46:31.8793 and declination $-$02:12:45.595. This is (63, 9) mas shifted from the nominal position, or $\sim$1/3 of the synthesised beam, probably mostly due to phase errors in transferring solutions from the phase-reference source. This is within the expected ALMA astrometric accuracy if no special measures are taken. The proper motion uncertainty contributes $\sim$12 mas to the position uncertainty at our epoch.

\subsection{Combination of compact and extended array data}

We channel-averaged the ACA data to match the spectral resolution used for the continuum data TM configurations, and flagged the line channels in the averaged data. We combined continuum-subtracted high spectral resolution data sets  for each spectral window separately for the three configurations. We compared visibility amplitudes on overlapping baseline lengths. The more extended spectral lines showed a small (a few percent) excess on ACA baselines, probably due to flux resolved out by TM configurations. Nevertheless, all were found to be consistent (within the scatter). The intrinsic sensitivities were used as weights of the individual data sets, determined by the antenna area, integration time per sample, and channel width (the last being the same for all data sets), modified by the inverse variance of calibration solutions. These were in the ratio of approximately 1:5:6 for the ACA:TM2:TM1 data. These native data weights give the highest sensitivity for a combined image. Finally, we made images from the combined ACA, TM1, and TM2 data for the continuum and each spectral line. Image cubes were made applying the primary beam correction. For compact emission, where the correction was negligible, uncorrected cubes were analysed. The largest angular scale that is reliably imaged is 15'', emission which is smooth on scales just over 15'' will still leave artefacts.

\section{ALMA data description} \label{datadescription}

In this section we give a detailed overview of all the morphologies identified in the acquired ALMA data. We analyse only the combined data cubes as they have the best combination of sensitivity and resolution. We begin by focusing on the continuum emission. We then proceed to the molecular emission, discussing the spatial emission distribution of SO$_{\rm 2}$, SiO, and finally CO.

\subsection{Continuum emission} \label{continuum}

The continuum emission of EP Aqr is shown in the left panel of Fig. \ref{cont}, the cumulative flux distribution in the right panel. The continuum peak is 17 mJy/beam, with a total flux within the 3$\sigma$ contour of 1.33 Jy. The diameter of the 3$\sigma$ contour is approximately 0.5''. Spatially, the emission is primarily centrally condensed, with no large deviations from spherical symmetry. Faint signatures in the contours tracing the 3 and 5 times noise rms exhibit some degrees of deviation from spherical symmetry. In particular, two filamentous features are seen to extend away from the stellar position to the south-south-east and south-south-west. These structures extend beyond 0.5'' away from the stellar position, which translates to an absolute distance greater than 65 AU.

Assuming the star has a brightness of 4800 $\lso$ and a temperature of 3236 K \citep{Winters2007}, we can estimate the contribution of the stellar flux to the continuum (by assuming black-body emission in a bandwidth 1.6 GHz around the rest frequency of 230.5 GHz) to be 2.27$\times$10$^{\rm -2}$ Jy, or approximately only 1.7\%\ of the flux in the 3$\sigma_{\rm rms}$ contour. This is an upper limit assuming that the dust is optically thin. \citet{Groenewegen1997} calculated that free-free emission is negligible below 3.3 mm, so it does not contribute to the observed continuum. The residual $\sim$1.30 Jy can thus be reasonably attributed to dust emission. For a typical oxygen-rich dust opacity of 7 $\rm cm^2/g$ \citep{Demyk2017} and an estimated mean dust temperature of $\sim$700K (assuming a radial temperature power-law decay with index $-$0.5) in the 3$\sigma_{\rm rms}$ region, we calculate the dust mass to be $\sim$5$\times$10$^{\rm -6} \mso$, assuming a mass-loss rate of $\sim$1.2$\times$10$^{\rm -7}\,\msoy$. This is an unusually large amount of dust;  for a typical dust-to-gas mass ratio of 2$\times$10$^{\rm -2}$ this would imply the equivalent of 2500 years of mass loss. This either means that the wind is exceptionally slow within the 3$\sigma$ contour (having a radial velocity component of only $\sim$80 m/s), or that the inner wind is dominated by a mechanism that impedes the matter ejected by the AGB star from escaping the inner 1''. We elaborate on this idea in Sect. \ref{SO2discus}.

\begin{figure}[]
        \centering
        \includegraphics[width=8.5cm]{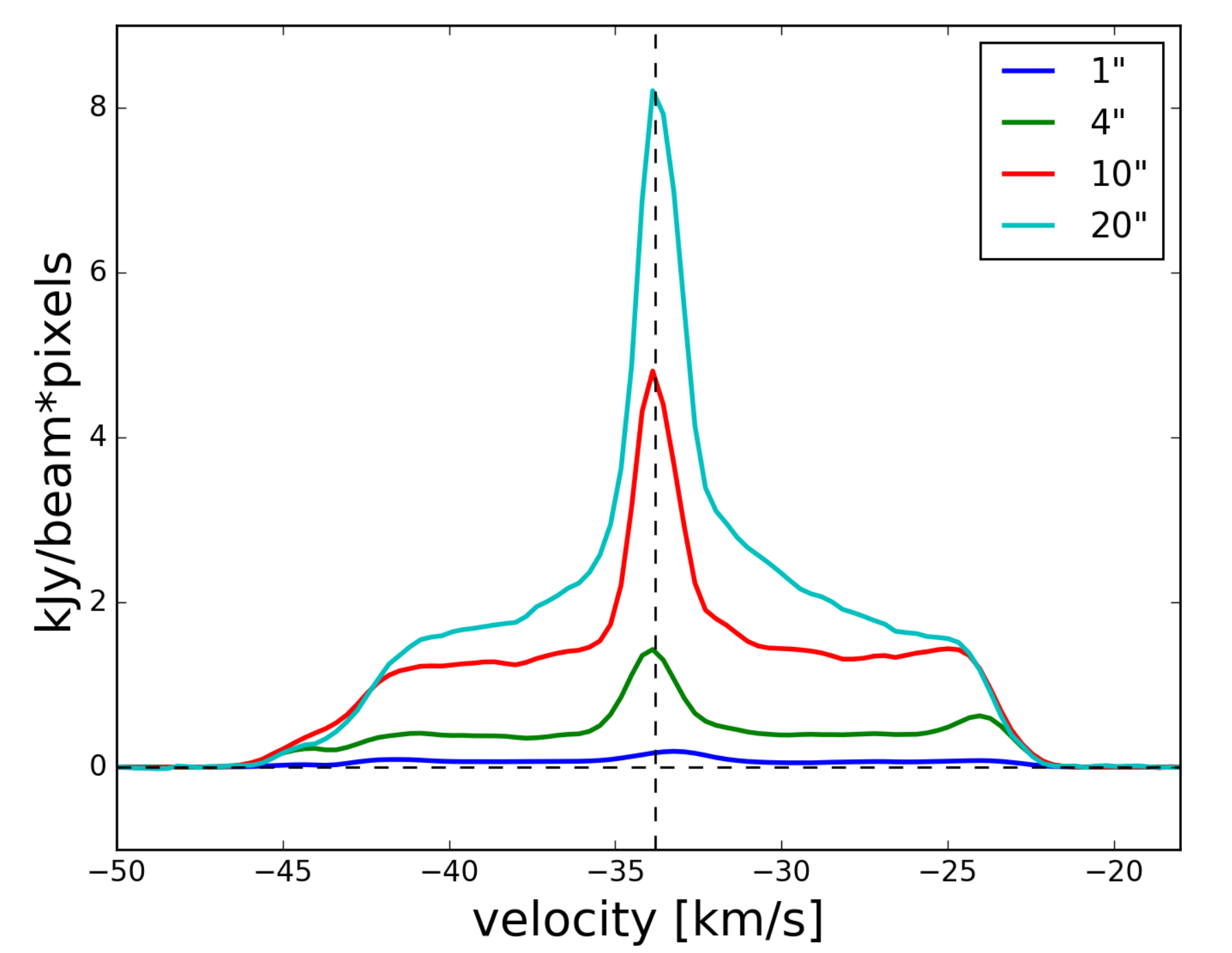}
        \caption{Spectral line of the CO emission for different circular apertures (diameters in legend). The shape is explicitly dual in nature with a broad plateau superposed by a bright narrow central peak. Black dashed lines have been added to the plot to indicate the zero flux level and the $v_{*}$.
        \label{COline}} 
\end{figure}

\begin{figure*}[]
        \centering
        \includegraphics[width=17cm]{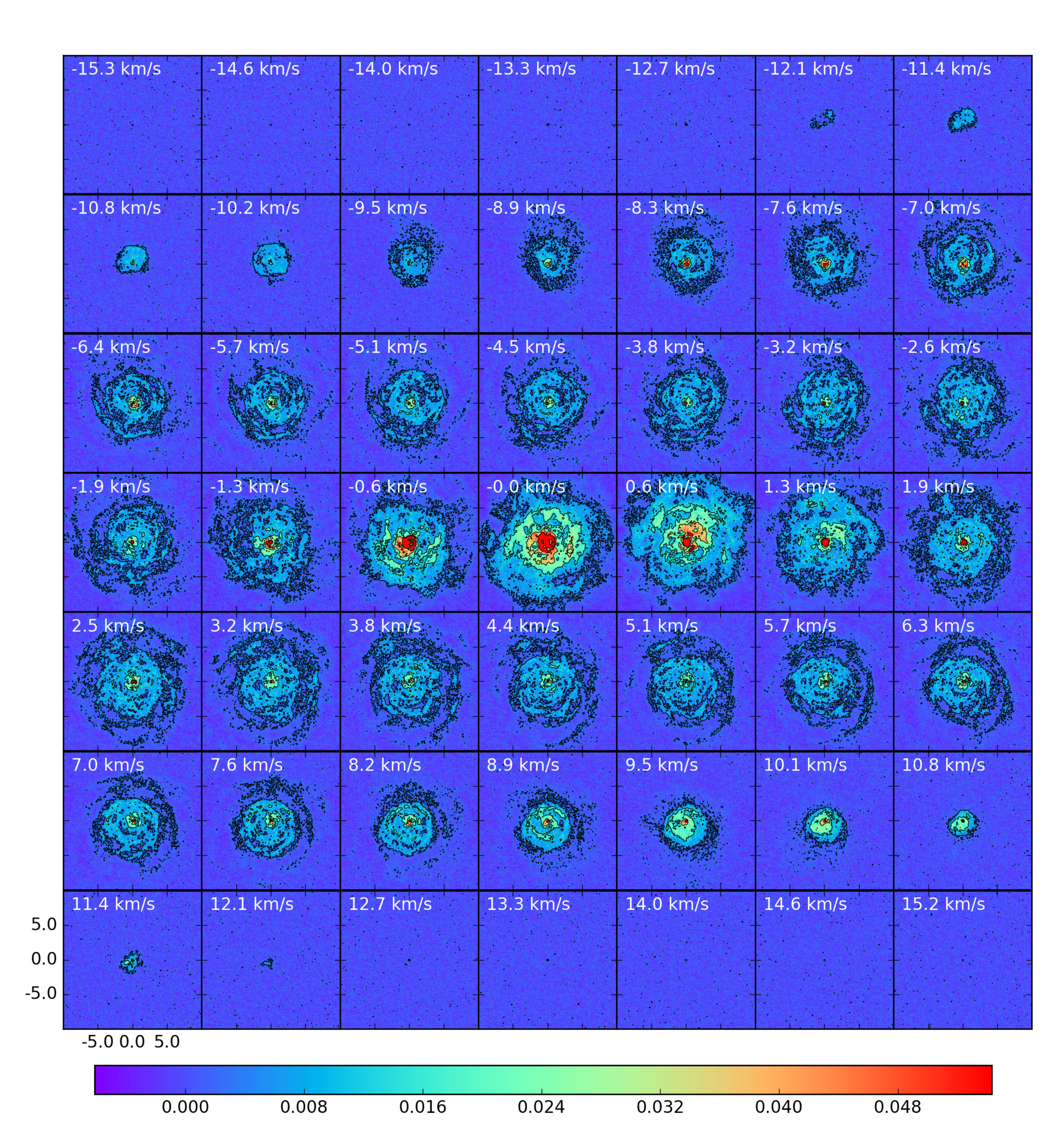}
        \caption{Continuum-subtracted channel maps of the CO emission corrected for $v_{*}$. The central observed frequency is the average over the observation epochs. The contours are drawn at 3, 12, 25, and 40 times the rms noise value outside the line (1.25$\times {\rm 10}^{\rm -3}$ Jy/beam). Length scales are indicated in the bottom left panel. Flux units are in Jy/beam. The continuum peak is located in the centre of the brightest contour. The beam size is not illustrated on the plots because it is too small (0.184''$\times$0.171''). The emission exhibits a high degree of structure. Faint filaments are visible at large distances from the continuum peak position. At $v$=0$\,\kms$ a spiral shape can be seen, with a bright centre.
        \label{COchancont}} 
\end{figure*}

\subsection{Molecular emission}

Three molecular lines were detected using the spectral setup delineated in Table~\ref{spw}. Spectral window 0 contained only the rotational transition $^{\rm 12}$CO(v=0, $J$=2$-$1); spw 1 contained the emission of SO$_{\rm 2}$(v=0, $\rm{28}_{\rm 3,25}-\rm{28}_{\rm 2,26}$), and yielded a non-detection for NaCl(v=0, $J$=18$-$17); spw 2 detected the rotational transition $^{\rm 28}$SiO(v=0, $J$=5$-$4); spw 3  contained only continuum emission, leaving SiS undetected in the wind. This last is not unexpected for low mass-loss rate winds of M-type AGB stars (Danilovich et al. \emph{in prep.}, Decin et al. \emph{in prep.}). A summary of the detected lines is presented in Table~\ref{idline}. The high spatial resolution of the most extended antenna configuration, in combination with the augmented maximum recoverable scale ensured by the additional two configurations yields  an exceptionally detailed and complete view into the morphological complexity of the circumstellar environment of EP Aqr. In this section we present the complexity of the spatial emission distribution through the channel maps, and the spectral nature of the emission through the spectral lines. The continuum-subtracted channel maps (with contours) of the detected CO, SiO, and SO$_{\rm 2}$ transitions are shown in Figs. \ref{COchancont}, \ref{SiOchancont}, and \ref{SO2chancont}, respectively, and the spectral lines in Figs. \ref{COline} and \ref{SiOline}, respectively. We present the Doppler velocity of spectral features in the channel maps as ($v_{\rm lsr}-v_{*}$), i.e. such that the stellar velocity $V_{\rm sys}=0\,\kms$.

\begin{figure*}[]
%         \centering
        \sidecaption
        \includegraphics[width=12cm]{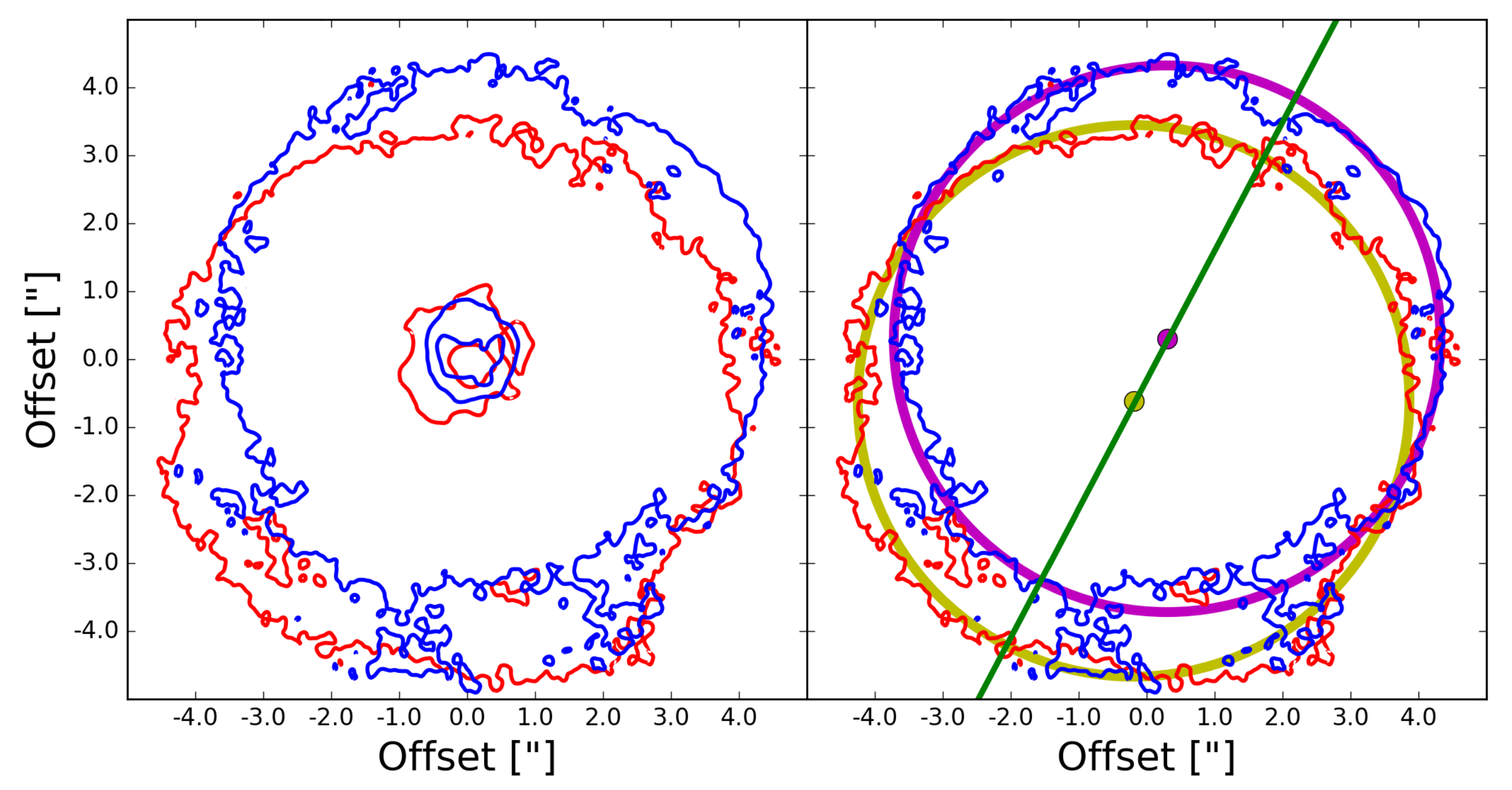}
        \caption{\emph{Left panel:} Stereogram of the CO cube. The blue (red) contours correspond to the velocity-averaged emission profile of the blue-shifted (red-shifted) portion of line outside of the central peak. The contours are drawn at 3, 13, and 23 times the rms noise value outside the line (1.25$\times {\rm 10}^{\rm -3}$ Jy/beam). \emph{Right panel:} The purple and yellow circles show the circular nature of the 3$\sigma_{\rm rms}$ emission. The purple and yellow dots represent the centres of the respective circles. The green line (150$^\circ$  from north, anticlockwise) is the axis along which these circle centres are separated, and therefore the axis along which the system is inclined.
        \label{stereo}} 
\end{figure*}

\begin{figure*}[]
        \centering
        \includegraphics[width=17cm]{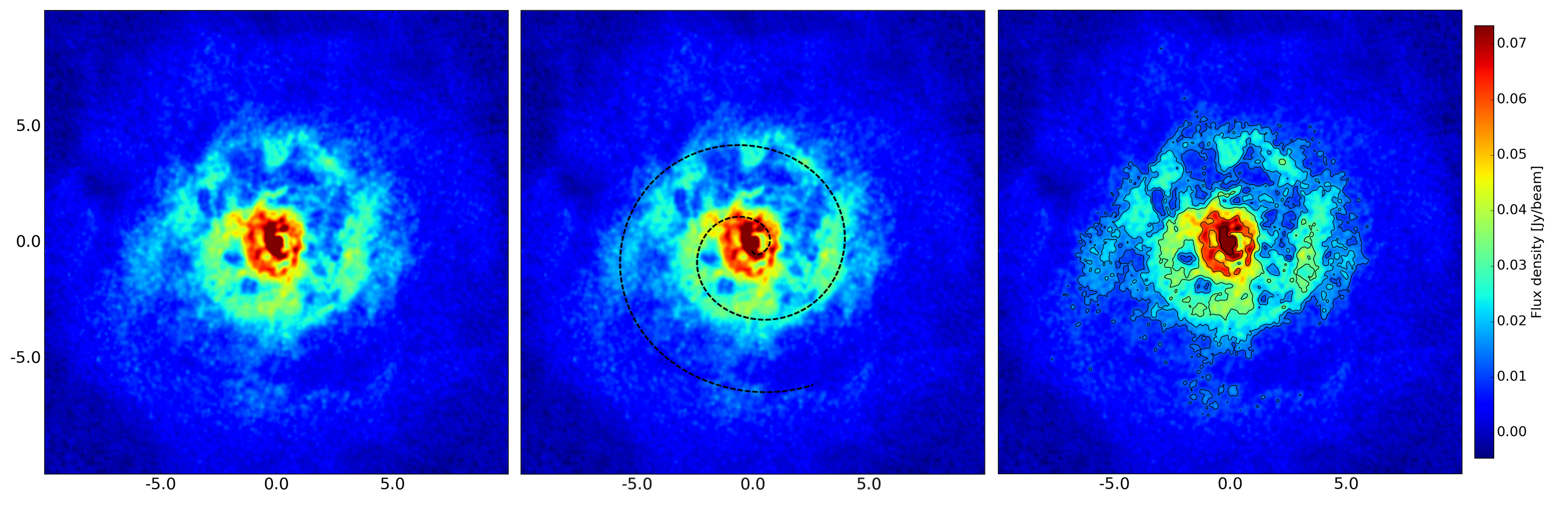}
        \caption{\emph{Left panel:} Central channel (at $v=0\,\kms$) of the CO emission. The spiral shape is clearly visible. \emph{Middle panel:} A black dashed Archimedean spiral is plotted over the channel's emission distribution, to guide the eye. \emph{Right panel:} Contour map of the central channel. Contours are drawn at 10, 15, 25, 45, and 65 times the rms noise value outside the line (1.25$\times {\rm 10}^{\rm -3}$ Jy/beam). The spiral exhibits a high degree of substructure, with voids and clumps. The central region of the spiral is bright and shows complexity beyond the expected spiral pattern.
        \label{COspir}} 
\end{figure*}

\begin{figure*}[]
%         \sidecaption
        \centering
        \includegraphics[width=8.5cm]{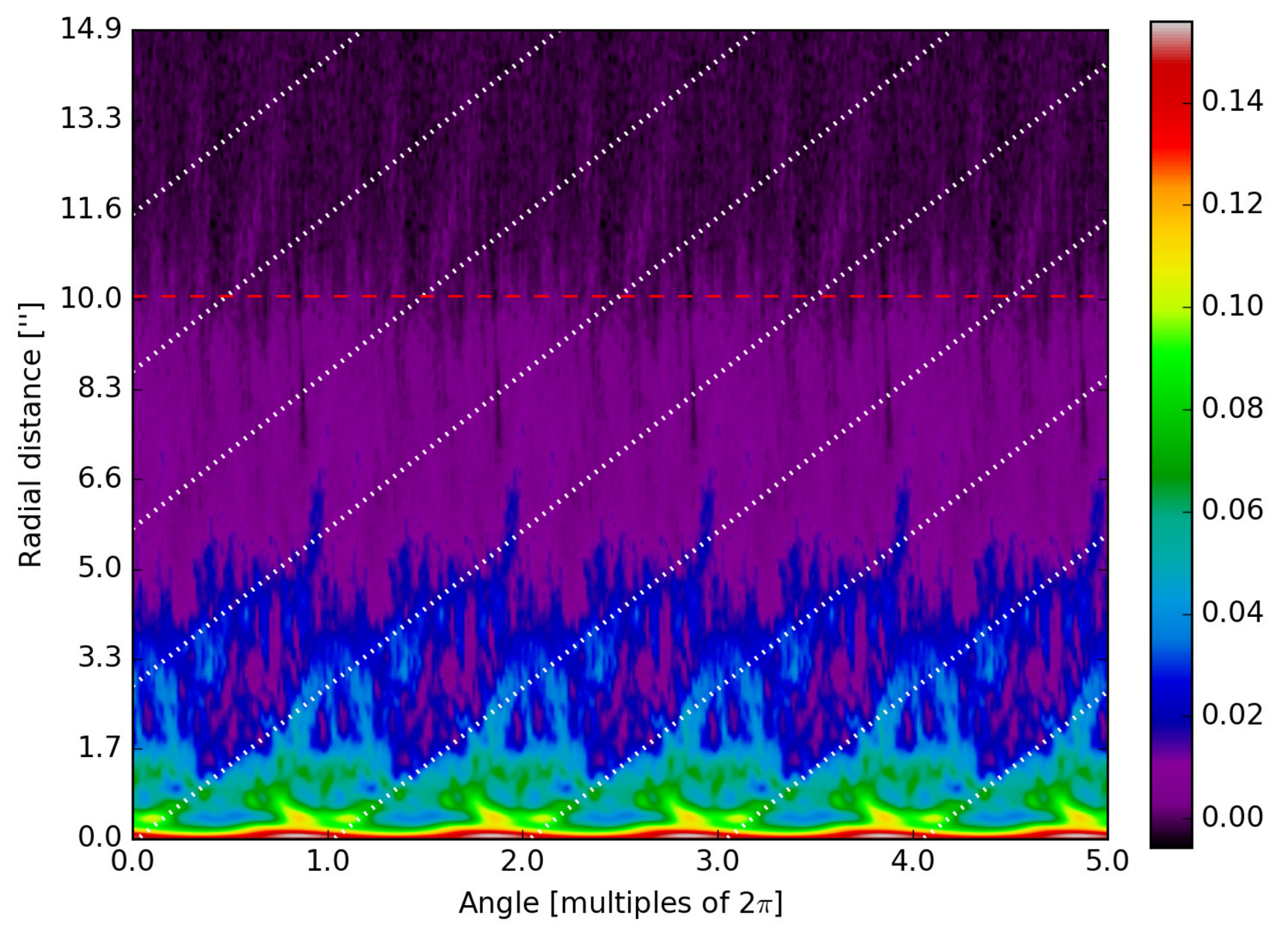}
        \includegraphics[width=8.2cm]{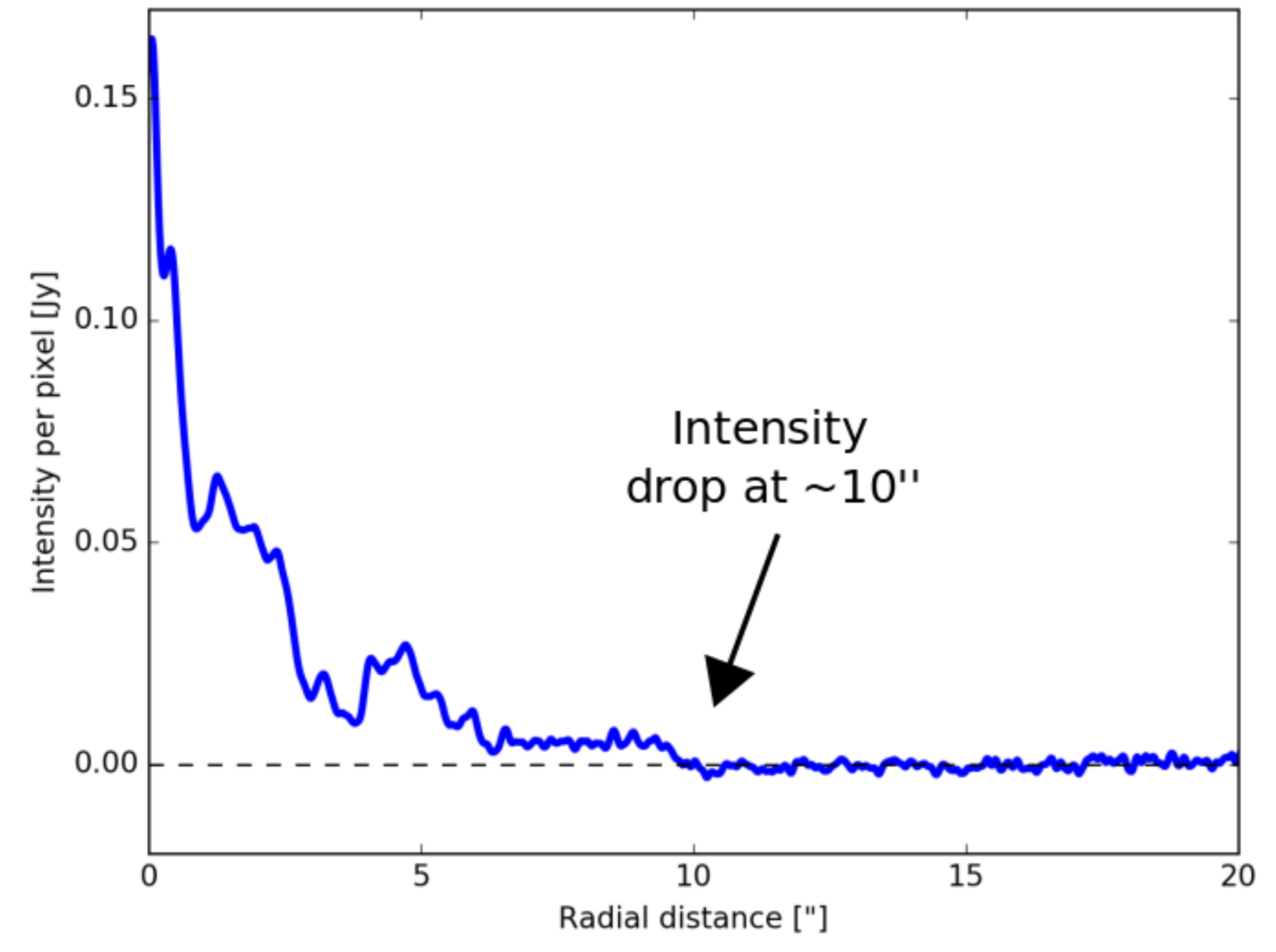}
        \caption{\emph{Left panel}: Radial intensity profile as a function of anticlockwise angle through the centre of the channel at $v=0\,\kms$, starting at a position angle of 150$^\circ$ (anticlockwise from north). Flux units are in Jy/beam. The spiral arm is visible as the radially expanding blue filament. White dotted lines have been overplotted to indicate the trend an Archimedean spiral would follow with a spiral arm separation of $\sim$1'' (or 135 AU at a distance of 135 pc). The spiral signal seems not to propagate into the inner regions (< 1.5''), and dilutes around 5''. At around 10'' a relatively sharp edge is seen in the signal, in approximately every radial direction. This is indicated by the red dashed line. \emph{Right panel}: Example of a radial intensity profile (at a position angle of 60$^\circ$) to better show the intensity drop at $\sim$10''.
        \label{COradial}} 
\end{figure*}

\subsubsection{$^{\rm 12}$CO $J=\ $3$-$2 emission} \label{COobs}

The carbon monoxide spectral line, shown in Fig. \ref{COline}, shows the peculiar dual nature identified and investigated by \citet{Winters2007}. It possesses a broad emission feature defining the width of the line ($\sim$23$\,\kms$), superposed by a bright narrow central peak ($\sim$3$\,\kms$). The shape of the emission at offset positions measured by \citet{Winters2007} shows that the source of this peculiarity is extended. This is also reflected in the scaling of every feature in the ALMA spectral line with increasing aperture size, as shown in Fig. \ref{COline}. Work by Homan et al. (2015, 2016) has shown that the peculiar dual nature of the CO line can be caused by an equatorial density enhancement (EDE) embedded in a radial outflow. Either the spectral profile testifies to the presence of a nearly face-on spiral structure that is geometrically confined to a very limited height or it originates from a relatively extended disk-like EDE, oriented face-on, subject to predominantly tangential rotation, and embedded in a bi-conical outflow.

Figure \ref{COchancont} presents the channel maps of the CO emission. The broad plateau in the spectral line originates from what seems to be a large-scale, rather inhomogeneous outflow, with a maximal velocity of $\sim$12$\,\kms$. The inhomogeneities manifest most prominently as small-scale clumps and larger-scale filaments of emission. These filaments have the tendency to trace semi-concentric shells and shell-segments, as can be seen in the channels from about $-$9$\,\kms$ to $-$3$\,\kms$ and from about 3$\,\kms$ to 8$\kms$. 

The predominantly spherical nature of the features in the wind suggest that the line plateau traces a centrally condensed radial velocity field. However, the concentricity of these filaments is dependent on the measured Doppler speed of the gas along the line of sight. This can be seen by comparing the emission in the $-$7$\,\kms$ channel with the emission in the 7$\,\kms$ channel. The overall circular emission in the $-$7$\,\kms$ channel has a tendency to move towards the north-west, while the emission in its 7$\,\kms$ counterpart pulls more towards the south-east. Assuming a radial outflow, this small offset between the positions of the virtual centres of the concentric shell-like patterns should originate from a minimal inclination, which would geometrically separate the projected centres of both the blue-shifted and red-shifted hemispheres. This separation thus permits us to deduce the axis along which the system is inclined. To find this axis we turn to the use of stereograms, which are velocity-averaged intensity maps. To limit the analysis to the portion of the line outside the central peak, we integrate from $-$14 to $-$2$\,\kms$ (shown as the blue contours) and from 2 to 14$\,\kms$ (shown as the red contours). The resulting stereogram is shown in the left panel of Fig. \ref{stereo}. The outermost contours have been fitted (by eye) with the yellow and purple circles in the right panel. The circle centres are indicated with respectively coloured dots. The green line represents the axis along which these two centres are separated, and is therefore the axis along which the system is inclined. We define the angle that the green axis makes with the horizontal axis as being the position angle of the system, which takes a value of $\sim$150$^\circ$, measured anticlockwise from north.

The narrow central emission peak is confined to the $-$1.3$\,\kms$ to 1.3$\,\kms$ portion of velocity-space. It manifests in the channel maps as a strong emission enhancement compared to the mean values in the plateau. Zooming in on the central channel a clear nearly face-on spiral signal emerges from the data, as shown in Fig. \ref{COspir}. The spiral consists of approximately two complete windings. It has a measured spiral arm full width of approximately 1'', which is comparable with the inter-spiral arm spacing. The spiral appears to the south-east in the blue-shifted wing of the peak, and recedes to the north-west of the stellar position in the red-shifted wing of the peak. The observed spiral feature is quite different from the previously detected spirals in carbon-rich environments:

\begin{itemize}
 \item The similar \textbf{width of the spiral arms and inter-spiral arm distance} stands in contrast to most carbon-rich spirals. Observations of spirals in C-rich environments as well as hydrodynamical models show spirals whose arm width always remains smaller than the distance between the spiral windings \citep{Mastrodemos1999,Kim2011,Kim2012,Maercker2012}. This typically happens because of the geometrical size difference between the local wind--binary interaction zone, and the total wind volume at that distance. In other words, the portion of the wind perturbed by the orbiting companion is small compared to the unperturbed outflow at that radius.
 \item The \textbf{inter-spiral arm distance} fluctuates as a function of radius, causing local deviations from the overall Archimedean behaviour (see Fig. \ref{COspir}, central panel). This is also reflected in a slightly elongated appearance in the north-west to south-east direction. Deviation from Archimedean trends are best visualised with a radius-angle intensity plot (see Fig. \ref{COradial}). Here the white dotted lines trace the expected Archimedean trend for a spiral arm separation of $\sim$1''. Though the spiral signal (blue tendril) emanating outwards from the central zone follows the Archimedean trend, it seems to be undulating along it.
 \item The \textbf{emission} in the spiral decreases significantly as a function of radius. The spiral signature becomes undetectable after two windings, which is unusual in comparison with carbon-rich counterparts. Because these data are an antenna configuration composite, with baselines down to 9m, this cannot be attributed to missing flux.
 \item The signal is \textbf{very confined in Doppler velocity-space}. The full spectral width of the spiral signal is only 2.6$\,\kms$. Taking into account the complexity of the hydrodynamical feedback between the motion of the mass-losing star and the wind--binary interaction, typical AGB spirals are found to have a rather large vertical height \citep{Mastrodemos1999,Kim2011,Kim2012}, and thus to encompass a relatively large range of velocities. The spectral confinement we observe here must thus follow from a specific kind of interaction. We elaborate on this in Sect. \ref{COdiscus}. We exhibit all peak channels in Fig. \ref{COzpcont} in the Appendix.
 \item The spiral signature \textbf{cannot be traced unambiguously to the centremost regions}. At distances of $\sim$1'' from the continuum peak position (100 AU at a distance of 135 pc) the spiral arm reaches into a zone where we cannot trace its spiral behaviour any further. This may partly be caused by the limited spatial resolution of the observation.
 \item This \textbf{central region exhibits high morphological complexity}, including a bright clumpy ring with a diameter of 1.5'', and an innermost hook-shaped, extremely bright central feature. We show the complexity of this central zone in Fig. \ref{COzccont} in the Appendix. These features are reminiscent of the wind Roche-lobe overflow hydrodynamical models performed by \citet{Mohamed2012}. We elaborate on this in Sect. \ref{COdiscus}.
\end{itemize}

The spiral exhibits a high degree of hydrodynamical inhomogeneity. Though the spiralling trend can be followed up to the point where the spiral signal fades completely, the actual spiral morphology is highly perturbed, resulting in an extremely clumpy appearance. This seems to be a property in which most observed spirals deviate quite substantially from their theoretical counterparts \citep[e.g.][]{Maercker2012,Decin2015,Kim2015}. Only the face-on spiral in the wind of the AB star AFGL 3068 \citep{Kim2017} possesses the smoothness associated with clean wind--companion interaction through a circular orbit.

In order to better gauge the radial properties of the emission, we created the radius versus angle intensity plot shown in Fig. \ref{COradial}. This figure is composed of concatenated radial profiles for different angles, spanning five full revolutions around the centre of the central CO channel. This figure exhibits four primary radial intensity regimes: (i) the inner region (from 0 to approximately 1.5'') shows a lot of structure, with a bright and compact inner zone; (ii) going further outwards to about 5--6'' the spiral dominates the radial profile. As previously mentioned, the spiral signal undulates along the white-dotted Archimedean trend, and has a width comparable to the spacing between the spiral arms; (iii) still further out is a zone which is relatively smooth, without any notable features; and (iv)  the boundary at $\sim$10'' is finally reached (indicated with the red dashed line), where the emission seems to make a relatively steep drop, decreasing the local intensities to barely detectable values. This can also be seen in Fig. \ref{cutoff} in the Appendix, where the image shown in Fig. \ref{COspir} has been fully saturated artificially. We might argue that this cut-off appears because the diameter of this cut-off zone exceeds the maximum recoverable scale of the most compact configuration or because of the reduced primary beam sensitivity. However, the primary beam sensitivity drops to $\sim$1/2 at $\sim$19'' radius for the 7m data set, so the sensitivity effect will contribute at most $\sim$20\% to the flux drop. And though missing flux undoubtedly partly explains the low intensity values at radii larger than 10'', it cannot introduce sharp intensity drops in the interferometric data (see Fig. \ref{COradial}, right panel). We tested this by simulating a synthetic observation of a smooth outflow with the approximate properties of the wind of EP Aqr. Though the emission indeed levels down to zero when the angular size of the smooth emission exceeds the MRS, it does so smoothly (at a radius of $\sim$8.5'') and not abruptly as  we observe in the data. We therefore tentatively conclude that this cut-off may be real, and part of the observed morphology. We elaborate on this idea in Sect. \ref{COdiscus}.

\subsubsection{$^{\rm 28}$SiO $J=\ $5$-$4 emission} \label{SiOobs}

\begin{figure*}[]
        \centering
        \includegraphics[width=17cm]{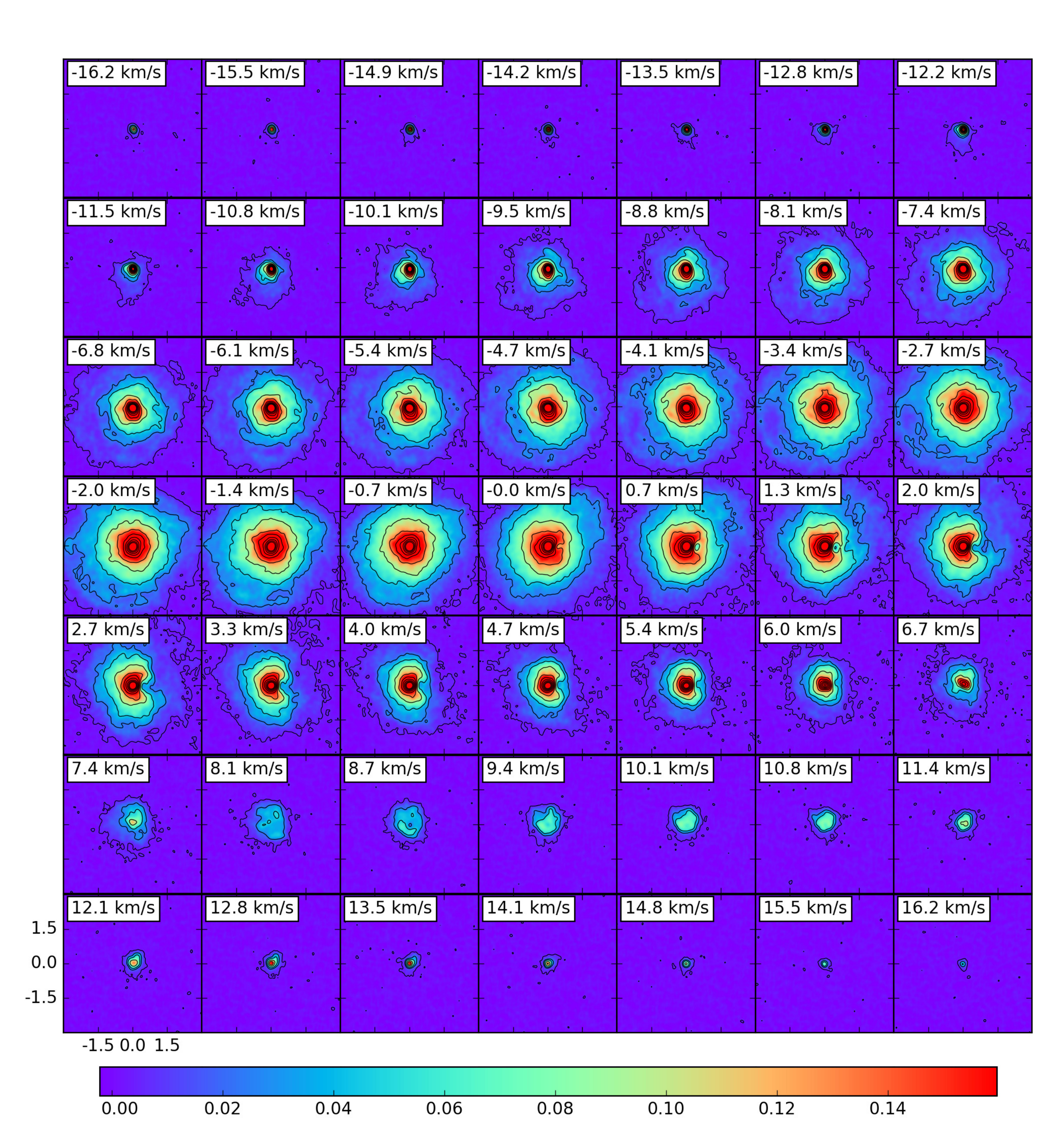}
        \caption{Continuum-subtracted channel maps of the SiO emission. The contours are drawn at 3, 20, 36, and then every multiple of 36 up to 291 times the rms noise value outside the line (1.2$\times {\rm 10}^{\rm -3}$ Jy/beam). Length scales are indicated in the bottom left panel. Flux units are in Jy/beam. The continuum peak is located in the centre of the brightest contour. The beam size is not illustrated on the plots because it is too small (0.205''$\times$0.187''). The emission is mostly spherical in nature, except for a distinct hole present to the west of the peak flux position, in the channels between $\sim$6 and $\sim$0$\,\kms$.
        \label{SiOchancont}} 
\end{figure*}

The $^{\rm 28}$SiO emission, shown in Fig. \ref{SiOchancont}, is more centrally confined than the CO emission, and is confined mostly to the central 4'' $\times$ 4'' portion of the complete field of view. On first sight it is clear that the emission distribution differs substantially from the CO maps. Overall, the behaviour is predominantly smooth and  spherical, which is reflected by the concentricity of the contours. The spherical nature of the emission is also reflected in the `classical' shape of the spectral line, shown in Fig. \ref{SiOline}. The line is, however, slightly asymmetric, with a decreased red wing.

Looking closely at the channel maps this diminished red wing originates from a small patch $\sim$0.5'' west of the continuum peak position where the emission is substantially lower than the local surroundings. This feature is confined to a well-defined portion of the red-shifted velocity-space, from around 0$\,\kms$ to around 6$\,\kms$. By comparing the minimal flux of the feature with the flux in its direct vicinity, we conclude that it is most prominently present at $\sim$1.4$\,\kms$, though the contrast is comparable up to velocities of $\sim$3$\,\kms$. At its minimum value, it has a flux density of 2.94 $\times$ 10$^{\rm -2}$ Jy/beam, or only 16\% of the flux density compared to nearby emission.

The presence of this local void seems to affect the CSE further out. All channels exhibiting this local void have a substantially decreased emission in the western portion of the wind. For a radiatively excited molecule like SiO, it seems that  the local void reflects a substantially affected mean radiation field. This dimmed western portion of the wind also coincides with the part of the wind that would, to some degree, be geometrically shielded from the emission of the central emission if the local void  contained some kind of radiation obstruction. Because we cannot reasonably attribute this spectrally confined local void to a morphologial origin inherent to the wind of a single star, we suggest that this feature may represent a local environment caused by the presence of a companion that has the ability to chemically affect its surroundings. We elaborate on this idea in Sect. \ref{SiOdiscus}.

\subsubsection{SO$_{\rm 2}$ $\rm{28}_{\rm 3,25}-\rm{28}_{\rm 2,26}$ emission} \label{SO2obs}

Presented in Fig. \ref{SO2chancont} is the observed SO$_{\rm 2}$ emission in the ALMA data. The spatial distribution of the emission is primarily centrally condensed, with the emission within the $\sim$12$\sigma$ contour only barely deviating from seemingly ideal spherical symmetry. The fainter emission extends up to only $\sim$0.5'' away from the continuum peak position, yet already shows some degree of complexity. As this region remains inside the location of the tentative companion deduced from the SiO emission, these irregularities should at least partly be attributed to the mass-losing star itself. They manifest most prominently as either clumpy structures or filamentous tendrils, and are more confined to the south, which seems to coincide with the behaviour of the continuum emission (Fig. \ref{cont}). The SO$_{\rm 2}$ emission shows that on length scales of only $\sim$0.2'' (corresponding to an absolute distance of $\sim$ 25 AU), the wind already turns clumpy and should thus be described as such \citep{VandeSande2018}.

\begin{figure*}[]
        \centering
        \includegraphics[width=8cm]{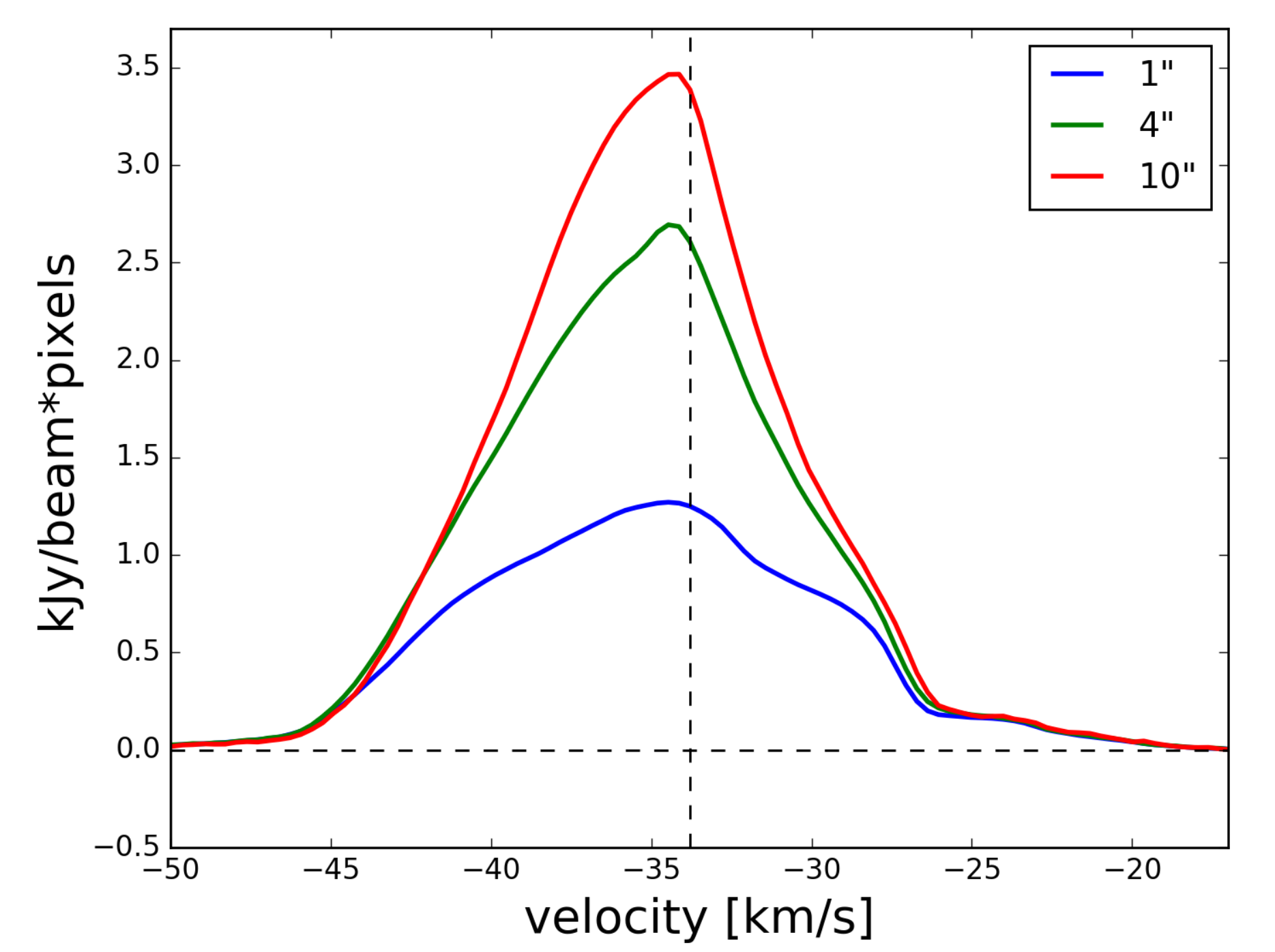}
        \includegraphics[width=8cm]{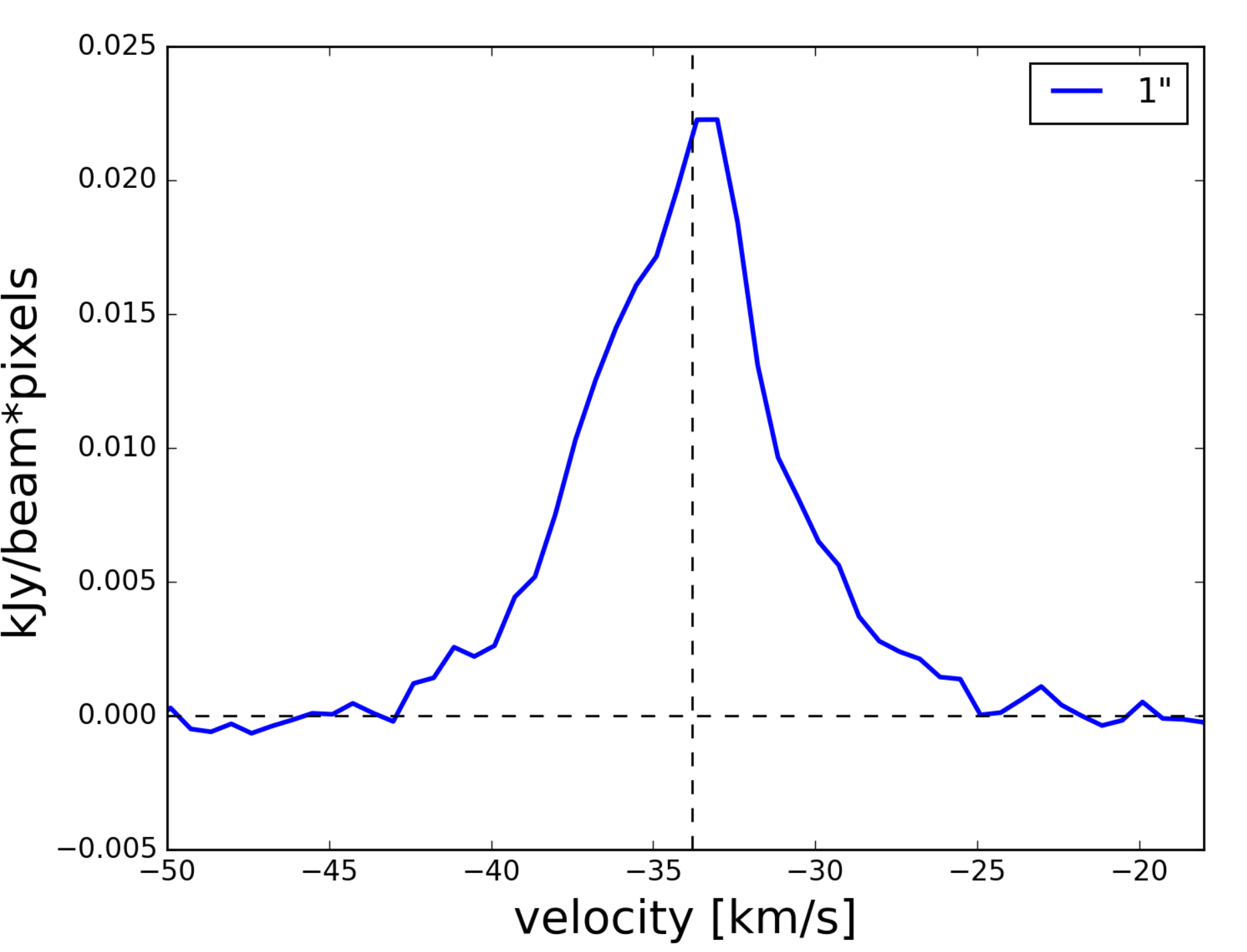}
        \caption{\emph{Left panel}: Spectral line of the SiO emission, for different circular apertures (diameters in legend). Its shape is predominantly triangular, with a substantially weaker red wing compared to the blue. Black dashed lines have been added to the plot to indicate the zero flux level and the $v_{*}$. \emph{Right panel}: Spectral line of the compact SO$_{\rm 2}$ emission, for a circular aperture with a 1'' diameter. Its shape is predominantly triangular. Black dashed lines have been added to the plot to indicate the zero flux level and the $v_{*}$.
        \label{SiOline}} 
\end{figure*}

\begin{figure*}[]
        \centering
%         \sidecaption
        \includegraphics[width=12cm]{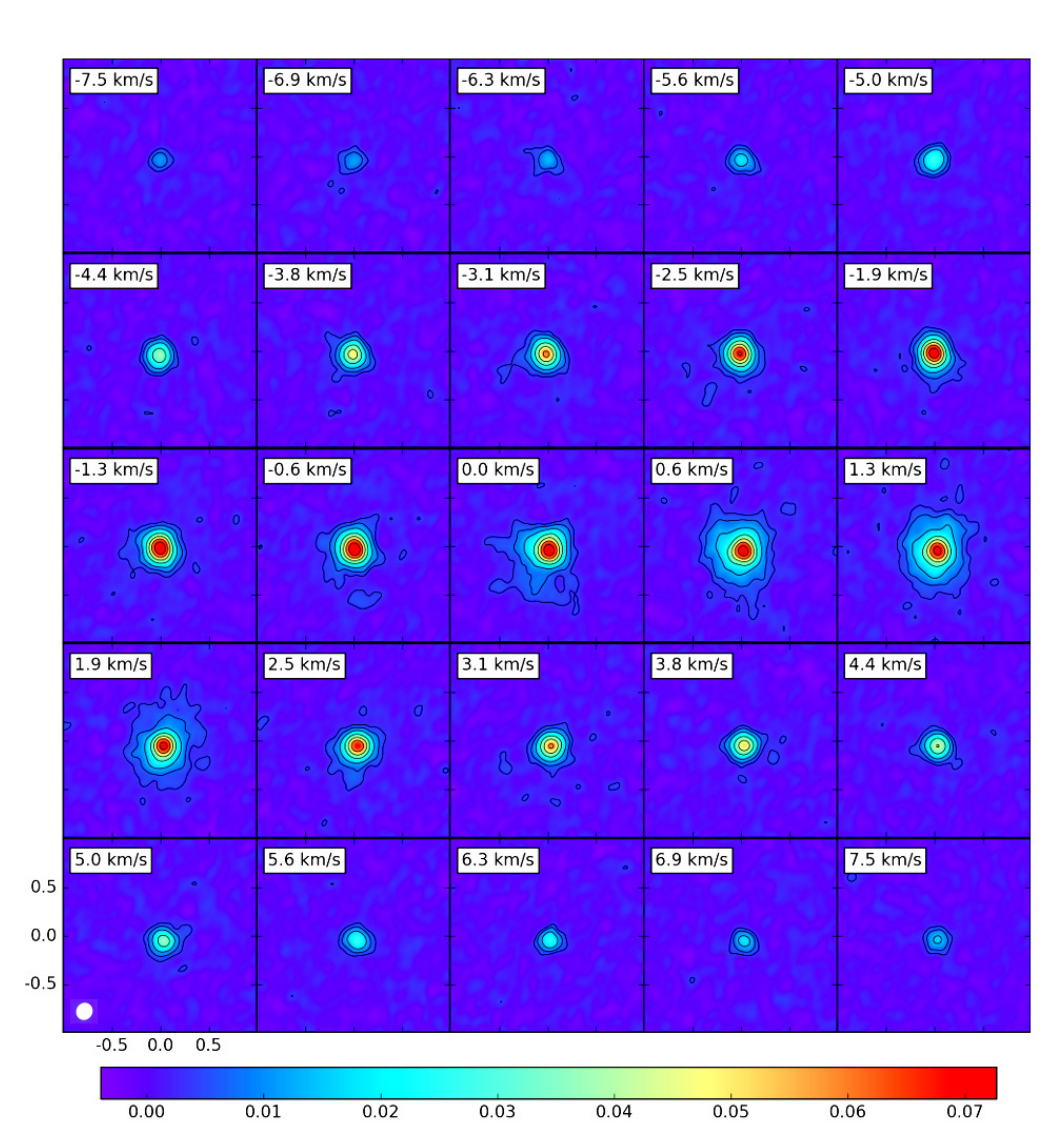}
        \caption{Continuum-subtracted channel maps of the SO$_{\rm 2}$ emission. The contours are drawn at 3, 6, 12, 24, 36, 48, and 60 times the rms noise value outside the line (1.18$\times {\rm 10}^{\rm -3}$ Jy/beam). Length scales and FWHM of the resolving beam are indicated in the bottom left panel. Flux units are in Jy/beam. The continuum peak is located in the centre of the brightest contour. Velocity labels are relative to the $v_{*}$. The emission is confined to the centremost regions of the wind, and is mostly spherical in nature.
        \label{SO2chancont}} 
\end{figure*}

\newpage

\begin{figure*}[]
%         \centering
        \sidecaption
        \includegraphics[width=0.7\textwidth]{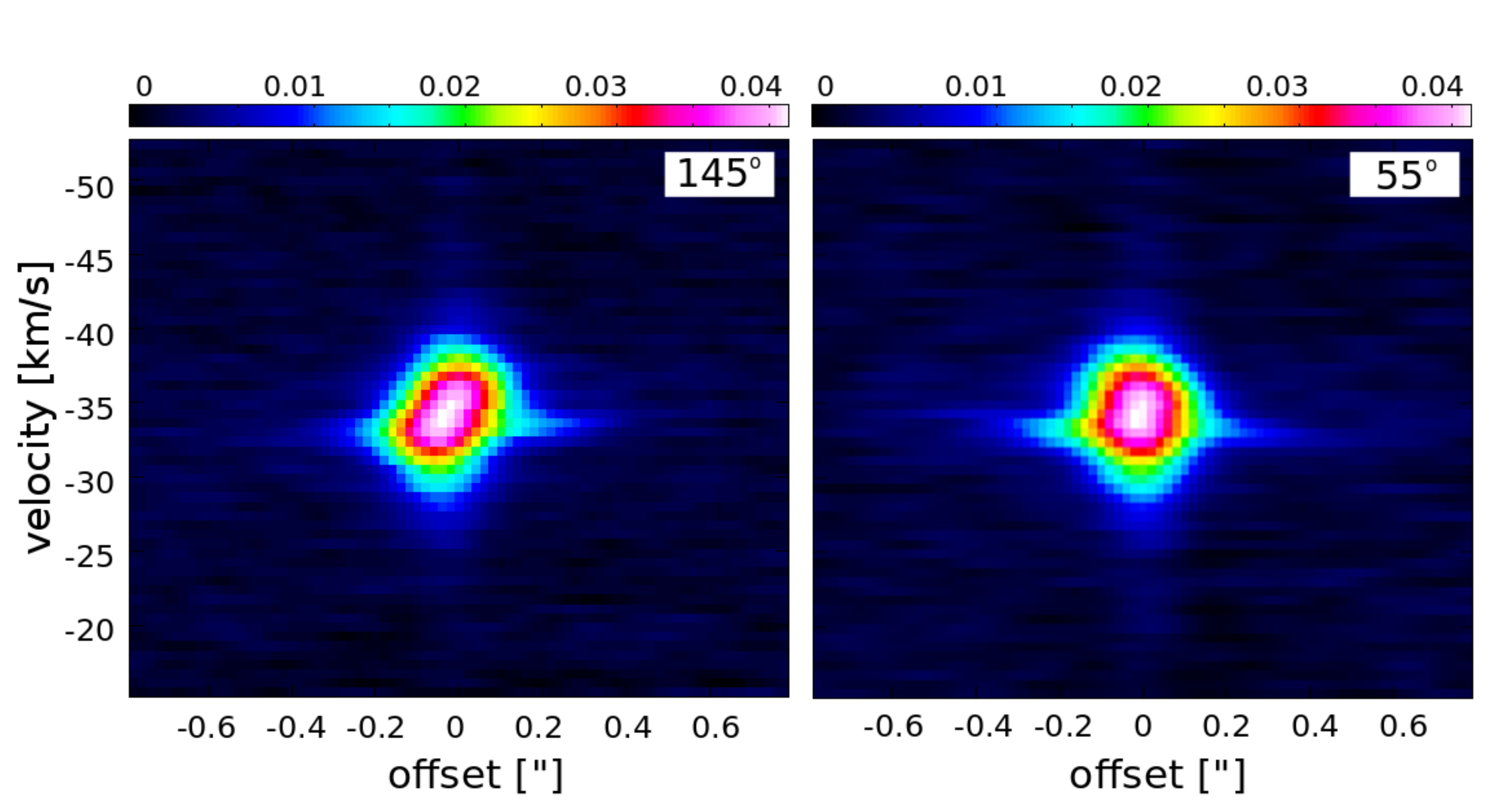}
        \caption{Orthogonal wide-slit PV diagrams of the SO$_{\rm 2}$ emission. Flux density in Jy/beam. The left panel shows the diagram generated by placing the slit at an angle of 145$^\circ$ (PV1), the right panel by placing the slit orthogonal to the first one (PV2). While PV2 shows a symmetrical emission distribution along both the central offset and velocity axes, the emission pattern in PV1 seems to be tilted clockwise. This is an indication of rotation around the 145$^\circ$ axis.
        \label{SO2pv}} 
\end{figure*}

\begin{figure}[]
        \centering
%         \sidecaption
        \includegraphics[width=7cm]{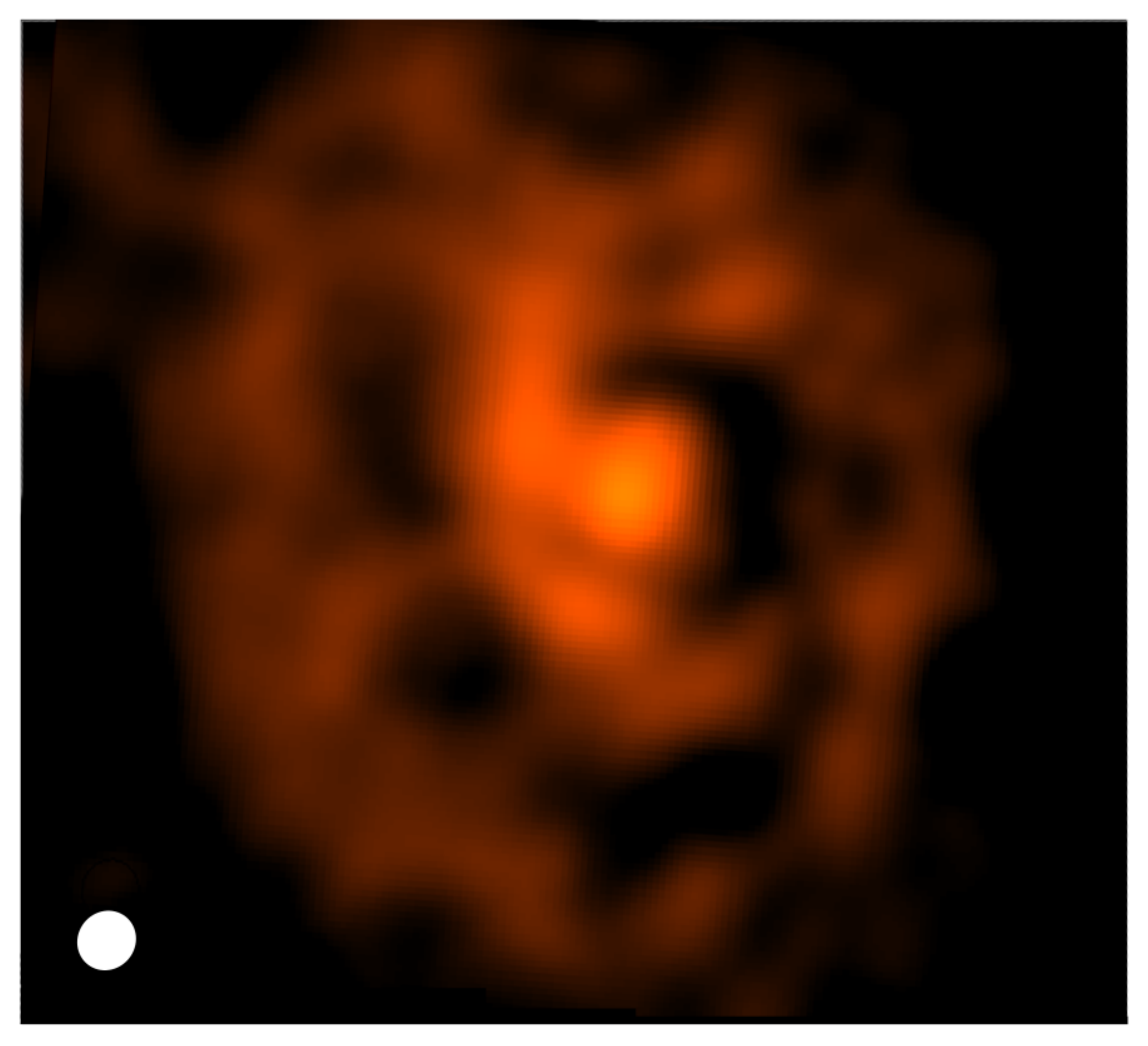}
        \caption{ Zoom in on the morphological complexity of the central region of the ALMA CO emission. Spatial scale is 2'', or 540 AU at a distance of 135 pc. Figure to be compared with Fig. 3 in \citet{Mohamed2012}.
        \label{wrlof}} 
\end{figure}

The SO$_{\rm 2}$ line is triangular in shape (see Fig. \ref{SiOline}), indicative that the material is probing the acceleration zone of the wind. The line, however, is slightly asymmetrical;  the blue side is slightly bulging while the red side is slightly hollowed-out.

The complexity of the innermost velocity field also subtly emerges from the channel maps. The distribution of the faint emission in the channels between $-$2.5 and 0.6$\,\kms$ reveals that the blue-shifted emission is somewhat elongated towards the south-east, while the red-shifted emission tends more towards the north-west. To  investigate this characteristic more closely, we have constructed position--velocity (PV) diagrams with axes at $5^\circ$ intervals through the centre of the image. We used a slit width of 1.5'' to maximise the signal-to-noise ratio, and ensure that all the emission is taken into account. Comparing these PV diagrams with their orthogonal counterparts we find that the highest degree of dissimilarity between the two diagrams appears at a chosen angle of 145$^\circ$ (and 55$^\circ$), as shown in Fig. \ref{SO2pv}. Here we see that the PV diagram along the 55$^\circ$ slit exhibits a compact, almost perfectly symmetrical signal, while the PV diagram along 145$^\circ$ shows a similar compact signal, but with a slight diagonal elongation, giving the appearance of a clockwise tilt. Combined, these signals point towards rotation, as shown by \citet{Homan2016}. We discuss this in more depth in Sect. \ref{SO2discus}.
\section{Discussion} \label{discus}

\subsection{Origin and nature of the spiral} \label{COdiscus}

The CO emission, tracking the bulk of the gas, reveals an almost face-on one-armed spiral, of which almost two full windings can be traced.  What could be the origin of this spiral structure?  As the majority of AGB stars are in binary systems, and perhaps all host planets, interaction between the outflow and a sufficiently massive and nearby companion may be the explanation of the observed CO morphology. The intricate emission features in the inner 2'' of the central CO emission maps is strongly reminiscent of hydrodynamical simulations of wind--binary interaction by  \citet{Mastrodemos1998} and  \citet{Mohamed2012}, where the latter authors performed tailored simulations for the Mira\,AB system in which the outflow of the AGB star Mira\,A is perturbed by the presence of its close companion Mira\,B. The wind--binary  interaction that ensues leads to what is known as wind Roche-lobe overflow (WRLOF), where the slow AGB wind is confined to the star's Roche lobe, while overflowing through the L1 Lagrange point. Gravitational interaction of the overflowing material with the companion produces an intricate feedback system where the stellar outflow material is ejected into the surrounding CSE through two distinct streams (through L2 and the stagnation point\footnote{In the co-rotating frame the stagnation point typically arises from the collision of two flows, in this case the Roche-lobe stream and the flow coming from the companion.}) which combine to form an annular stream. As this stream travels outwards, it creates the larger scale spiral observed in the wind. The morphology resulting from this particular type of wind--binary interaction is shown in Fig. 3 in \citet{Mohamed2012}. In Fig. \ref{wrlof} we show the emission pattern seen in the central regions of the CO channel at $v_{*}$. We compare this image with the bottom left panel of Fig. 3 in \citet{Mohamed2012}, an opacity map of the interaction zone. Though the two properties that are compared differ in nature, they likely still trace the same global morphological structure. Indeed, several of the predicted morphological features can be identified in the data of EP Aqr. The bright central region with a north and southward hook-like extension are strikingly similar, as are the eastern and western crescent-shaped `voids', the overall shape, and the morphological properties of the small-scale instabilities.

It must  be noted, however,  that the physical conditions in the simulation of \citet{Mohamed2012} differ from those in the inner wind of EP Aqr. Mira has a low-velocity wind ($\sim$5$\,\kms$) and a binary separation of $\sim$43\,AU (from a spatial separation of $\sim$0.47'' and an assumed distance of 92 pc \citeauthor{Vlemmings2015} \citeyear{Vlemmings2015}). The wind of EP Aqr is considerably faster ($\sim$12$\,\kms$) and the tentative companion at 68\,AU more distant, assuming that the SiO emission anomaly at $\sim$0.5'' is indicative of its location (see Sect.~\ref{SiOobs} and below). These differences may imply that WRLOF provides a plausible scenario for the mass-loss morphology of Mira, but not for EP Aqr. The physical set-up of the hydrodynamical models of \citet{Mastrodemos1998} are closer to the conditions in the inner wind of EP Aqr. Interestingly, they produce a morphology in the orbital plane which somewhat resembles the features seen in the inner parts
of the CO emission.

Can we place constraints on the velocity field in the almost face-on spiral structure? In the following, we consider two such global morphologies: \begin{itemize}                                                                                                                                                                   \item (a) a spherical wind with a strictly radial velocity structure with an embedded equatorial spiral morphology, characteristic of wind--binary interactions in C-rich AGB CSEs \citep{Kim2011,Kim2012};                                                                                                                                                                         \item (b) a strictly tangential velocity field, characteristic of differential rotation and a wide opening angle bi-conical outflow perpendicular to the midplane of rotation.                                                                                                                                                          \end{itemize}
In both scenarios, the equatorial structure is responsible for the CO peak emission near line centre and the bi-conical radial flow for the broad plateau-shaped line wings.

To explain the 1.5$\,\kms$ half width of the central part of the CO line in scenario (a), the spiral opening angle should be restricted to at most $\sim10^{\circ}$ to account for the estimated inclination angles. Such a  confined morphology is not in line with models of wind--binary spiral formation, which typically produce much wider structures \citep{Mastrodemos1999,Kim2011,Mohamed2012}. \citet{Mastrodemos1999} show the results of 18 different models, of which two (models M2 and M12) produce a spiral-shaped density enhancement which is vertically confined. However, even for these models the scale height of the spiral is at least double the value we measure here. Of course, the spiral of EP Aqr may be the geometrical manifestation of a set of physical conditions which have yet to be hydrodynamically modelled.

For this scenario (a) the width of the central peak allows us to constrain the inclination angle of the system (with respect to the face-on position of 0$^\circ$)  along the axis determined in Sect. \ref{COobs}. The maximum velocity of the outflow, given by the width of the CO line ($\sim$12$\,\kms$) would yield a projected velocity component of $\sim$1.5$\,\kms$ (half the peak width of 3$\,\kms$) under an inclination angle of $\sim$7$^\circ$. This value should be considered an upper limit. If the spiral gas suffered from small-scale stochastic motions, part of the central peak broadening of the CO line would be due to micro-turbulent broadening. If the micro-turbulent velocity amounted to $\sim$0.5$\,\kms$, the inclination angle would  only be $\sim$4$^\circ$.

As shown in Fig. \ref{COzpcont} in the Appendix, the highest projected velocities reside closer to the continuum peak position than the lowest projected velocities. This behaviour is consistent with our alternative model (b), that of a slightly tilted disk in tangential Keplerian-like rotation. The tangential nature of the velocity field does not permit the deduction of the vertical extent of the disk. For such a configuration it may also be more natural to expect an abrupt drop in the CO emission (seen at $\sim$10'') being the manifestation of an outer disk limit (see Fig.~\ref{cutoff}). In this case the observed spiral could be a hydrodynamical perturbation of (or in) the disk caused by the presence of a companion \citep{Goldreich1979,Ogilvie2002,Rafikov2002,Meru2017}.

The inclination angle of the orbital plane of the binary system hypothesised to cause this structure can be constrained if the Keplerian velocity of the companion can be deduced. In the wind--companion  interaction mechanism, this velocity is identified as the maximum tangential velocity of the gas. For a companion at an angular separation of $\sim$0.5'' or $\sim$65\,AU (see below) and a mass of EP\,Aqr of $\sim$1.5$\,\mso$ \citep{Dumm1998}, we find this Keplerian velocity to be $\sim$4.5$\,\kms$.  To explain the width of the peak, this would translate to an inclination angle of 18$^\circ$.  Also, assuming a micro-turbulent velocity component of $\sim$0.5$\,\kms$, the inclination angle would reduce to $\sim$12$^{\circ}$.

\begin{figure*}[]
%         \centering
        \sidecaption
        \includegraphics[width=17cm]{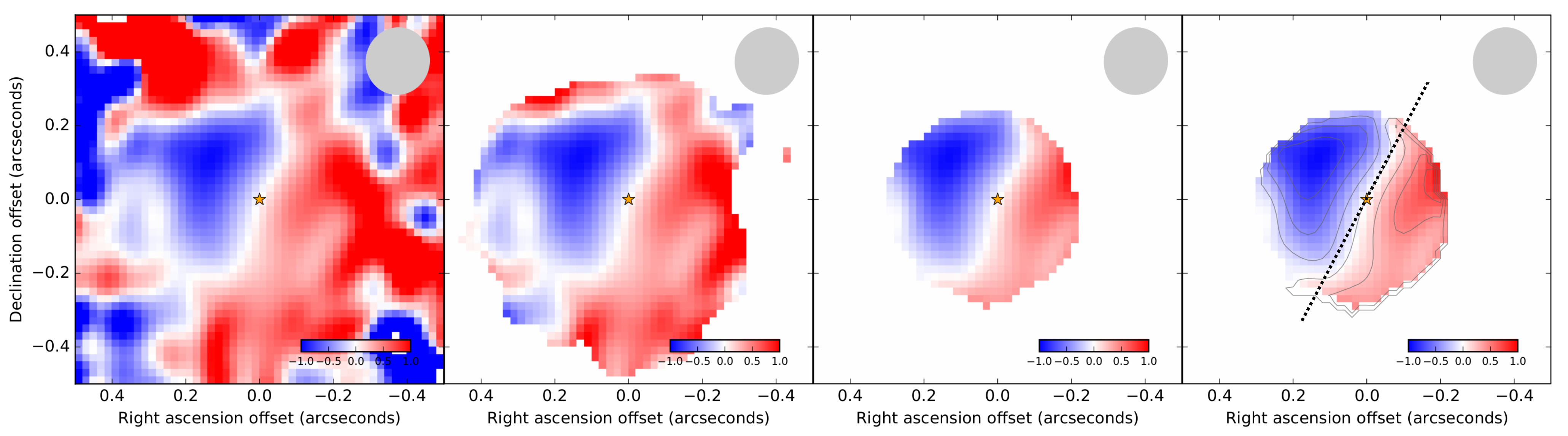}
        \caption{Moment 1 map of the central 1''$\times$1'' of the SO$_{\rm 2}$ emission. Colour scale is in$\,\kms$. The star is placed on the peak brightness position, velocities are measured with respect to this position. From left to right: (1) Complete moment 1 map; (2) Same as 1, but with an intensity cutoff at 0.002 Jy; (3) Same as 1, but with an intensity cutoff at 0.005 Jy; (4) Same as 3, but with contours at steps of 0.25$\,\kms$ away from zero. The black dotted line indicates the previously determined position angle of the system. The ALMA beam FWHM is shown in the top right corner.
        \label{mom1}} 
\end{figure*}

\begin{figure*}[]
%         \centering
        \sidecaption
        \includegraphics[width=12cm]{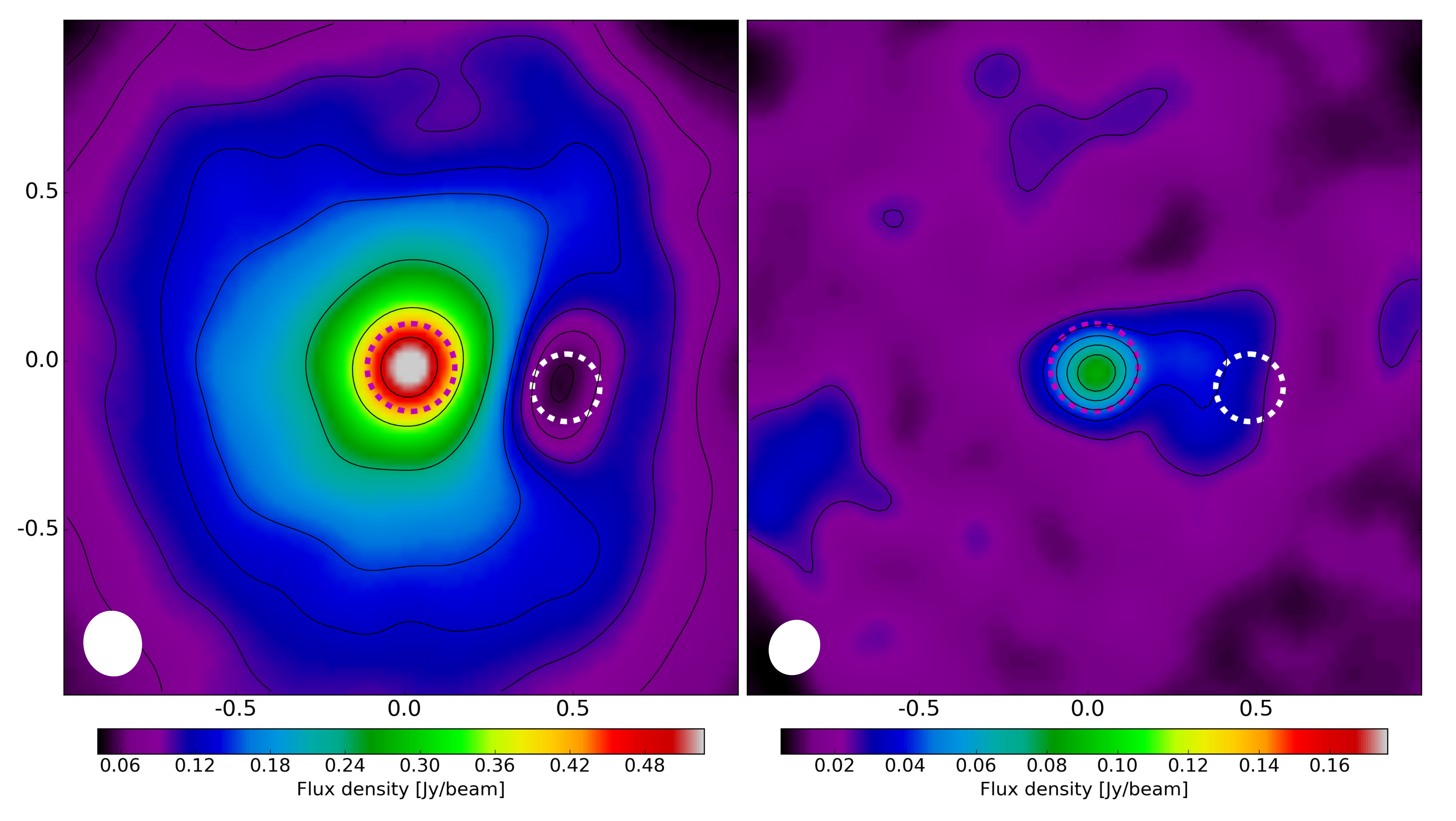}
        \caption{ \emph{Left panel:} SiO emission distribution in the 1.4$\,\kms$ channel, zoomed in on the central 2'' $\times$ 2''. The local void is clearly visible. The purple and white dashed circles have been added to guide the eye, centred on the continuum brightness peak location and the local void, respectively. \emph{Right panel:} CO emission in the same channel. A lane of gas is seen bridging the gap between the AGB star and the position of the local void.
        \label{hole}} 
\end{figure*}

\subsection{Inner gas rotation and deduction of the inclination of the rotation axis.} \label{SO2discus}

In Sect. \ref{SO2obs} we identified a relation between the SO$_{2}$ emission offset along the 145$^\circ$ angle (anticlockwise from north) and the projected velocity of this emission, visualised in the orthogonal wide-slit PV diagrams shown in Fig. \ref{SO2pv}. Such a dynamical behaviour is consistent with a rigidly rotating gas distribution \citep{Homan2016}.

The region probed by the SO$_{\rm 2}$ emission lies just within the orbital radius of the tentative companion. Hence, the emission is likely affected by dynamical effects induced by both the AGB star (e.g. surface pulsations) and the orbiting companion (e.g. gravitational drag). Hydrodynamical simulations by \citet{Mastrodemos1998,Mastrodemos1999} show that the dynamical complexity of the inter-binary zone can be substantial, and that the gas experiences a torque from the companion's motion resulting in a tangential velocity component, superposed onto the radial outflow. The magnitude of this torque-induced tangential velocity is largest if the masses of the host and the companion are not too different (at most a factor of 10) and can amount to a significant fraction of the wind speed at the orbital radius, though it never exceeds the wind speed.

To better understand the spatial distribution of the (projected) velocity field we have constructed a first moment map of the central SO$_{\rm 2}$ emission, displayed in Fig. \ref{mom1}. A clear rotation signature is seen in the SO$_{\rm 2}$ emission, with projected amplitudes up to 1$\,\kms$. The axis of rotation agrees well with the previously determined position angle of the system (see Sect. \ref{COobs}). The band of the highest projected velocities (to the north-west of the peak brightness position) can be expected to trace the equatorial plane of the rotation. Hence, this implies that the axis of rotation is inclined with respect to the plane of the sky with its south-eastern end pointing towards the observer, and its north-western end pointing away. This agrees with the spectral progression of the CO peak emission, where the blue-shifted portion of the peak tends more towards the south-east, and vice versa for the red-shifted portion of the peak. It also agrees with the relative positions of the `unperturbed' wind hemispheres visualised through the stereograms (Fig. \ref{stereo}).

Using the inclination angle of the system derived above (between 4$^{\circ}$ for a radial stream and 18$^{\circ}$ for a rotating disk), we can derive the actual rotation velocity of the gas. The highest measured projected velocities in the SO$_{\rm 2}$ correspond to an absolute velocity for the respective inclinations between $\sim$14$\,\kms$ and $\sim$3.5$\,\kms$. There are no currently known processes which would produce a torque of such magnitude on the sub-companion-orbit gas that the induced mean rotation speed exceeds the wind's previously deduced radial velocity of $\sim$11$\,\kms$. Typically, the tangential velocities induced by the companion's gravity remain below $\sim$20\% of the wind speed \citep{Mastrodemos1998,Mohamed2012}. Hence, higher inclinations are favoured, supporting the rotating disk hypothesis. This rotation may keep the outflow material confined to raddi smaller than the companion's orbit for a time much longer than the radial dynamical crossing time, possibly (partly) explaining the unusually large dust mass measured in this region (see Sect.\,\ref{continuum}).

\subsection{Origin of the local void observed in SiO} \label{SiOdiscus}

The local void in the SiO emission,  located $\sim$0.5'' (or $\sim$68 AU at a distance of 135 pc) west of the continuum peak position, could be the manifestation of a local environment, caused by the presence of a companion. The inclination orientation of the system, as deduced in Sect. \ref{SO2discus}, in combination with the westward position of the void indeed implies that this local environment resides in the red-shifted portion of the spectral line. We can estimate the size of this local environment based on the inclination of the system. Assuming a purely radial outflow (inclination $\sim$7$^\circ$), the velocity-width of the feature ($\sim$6$\,\kms$) can be explained by a local environment with a diameter of $\sim$20AU. For a tangential field inclined under an angle of $\sim$18$^\circ$, its size is more difficult to deduce, we cannot measure the local velocity field (gradient). However, assuming Keplerian dynamics and a mass of 1.5$\,\mso$ for the central star, and taking into account the position of the void with respect to the inclination axis, we can estimate the diameter of the void to be $\sim$30 AU.

Typical companions known to produce such local environments are white dwarfs (WDs), because they possess a strong ultraviolet radiation field. The presence of a WD is expected to produce substantical photochemical effects on the nearby gas of a stellar wind \citep{Ramstedt2014}. If not a WD, then we can place constraints on the properties of the companion based on its non-detection in the continuum. Attenuation of the companion's continuum flux by dust in the AGB outflow is likely to be modest. Assuming oxygen-rich dust with a mean opacity of 7 cm$^{2}$\,gr$^{-1}$ at 230.5 GHz, and a density power law index of -2, the radial optical depth amounts to 0.16, weakening the companion's continuum emission by only 15\%. Thus, if we assume that the companion is an unevolved main sequence star, then it can at most be an M6 dwarf if it is to remain below the ALMA detection limit at 230.5 GHz. Such objects have typical luminosities of $\sim$10$^{\rm -3} \lso$, typical effective temperatures of $\sim$2800 K, and masses of $\sim$0.1 $\mso$ \citep{Siess2000}. 

We also compare the emission distribution of the SiO channel at 1.4$\,\kms$ with the same CO channel, as seen in Fig. \ref{hole}. Where the local void (indicated with a white dashed circle) stands in clear contrast with the local surrounding medium and the continuum peak position (indicated with a purple dashed circle) in the SiO emission, the CO reveals a band of emission connecting the location of the AGB star with the location of the absorption feature. This bridging feature could be indicative of mass-transfer between the central star and the nearby object, or an increased production of CO at the position of the local void. \citet{Mastrodemos1998} predict that for certain physical set-ups efficient mass transfer happens from the AGB star to a companion, which is accreted by this companion through a compact accretion disk (too small to observe at the ALMA spatial resolution). This is in line with the hypothesis that the local void may indeed be caused by the presence of a companion.

\subsection{High-resolution observations reveal enhanced expansion velocities}

\begin{figure}[]
        \centering
        \includegraphics[width=8cm]{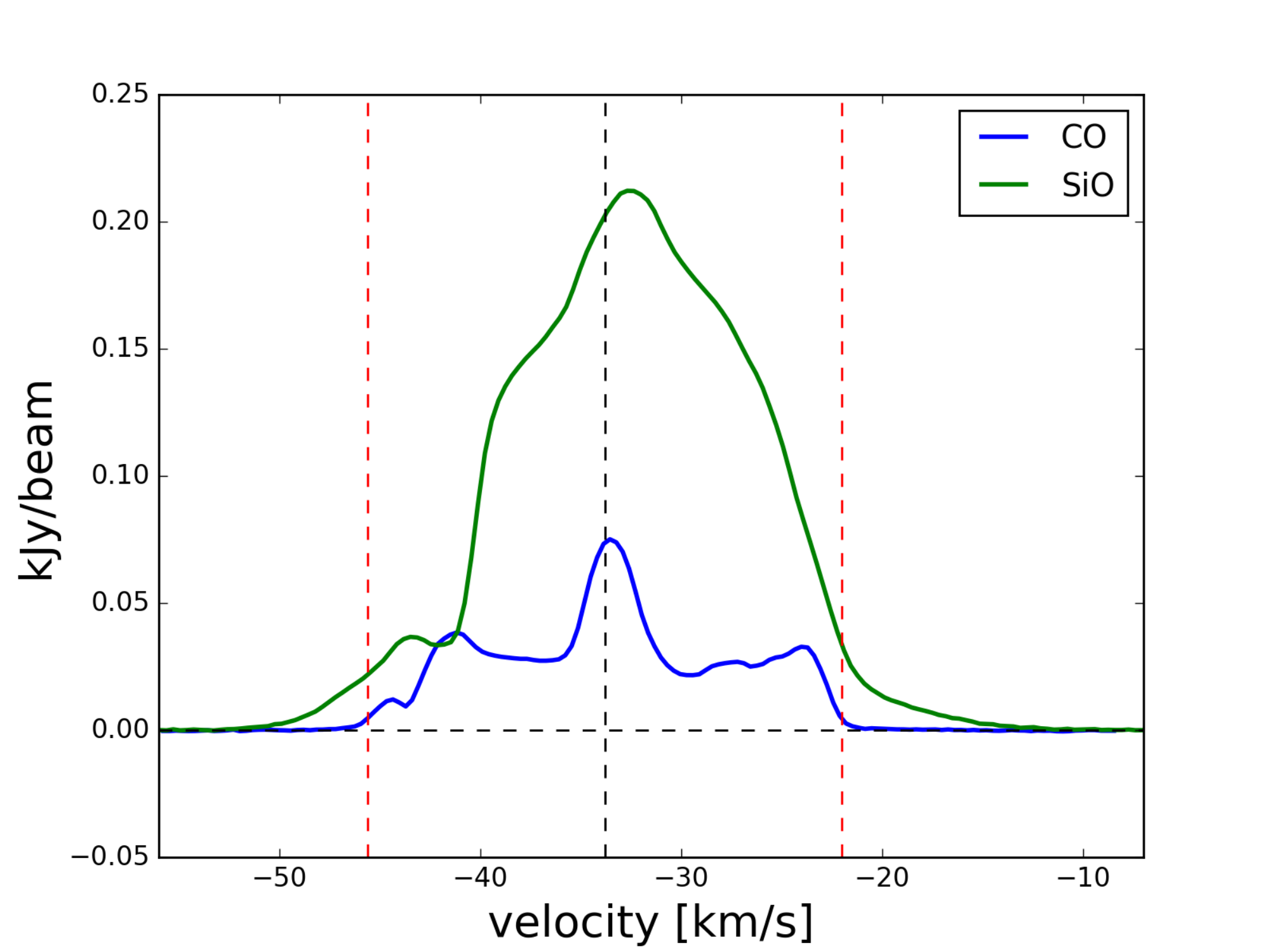}
        \caption{Comparison between the mean flux of the CO and SiO emission in a circular aperture of 1'', centre on the continuum peak. Black dashed lines have been added to the plot to indicate the zero flux level and the $v_{*}$. The SiO is intrinsically much brighter, and as a consequence it exhibits much broader high-velocity wings than CO. This is illustrated by the red dashed lines, which mark the velocities at which the CO line subsides below the noise.
        \label{wing}} 
\end{figure}

The CO and SiO lines, plotted over each other in Fig.~\ref{wing}, differ not only in line shape and strength but also in maximum width of the line wings. The CO data channels exhibit emission up to a maximum of $\sim$12$\,\kms$; the SiO signal exhibits a signal up to almost 20$\,\kms$ from line centre. The latter is much higher than any previous wind speed measrements. For both species, the highest velocities are concentrated along lines of sight close to the position of the AGB star, consistent with a somewhat bi-conical or bi-polar morphology of the AGB wind in which the highest velocities are reached in the polar direction \citep[see][]{Nhung2015}. This strong discrepancy between the widths of spectral lines of highly spectrally and spatially resolved AGB winds is also observed in the AGB stars IK Tau and R Dor (Decin et al. \emph{in prep}).

The main reason for the discrepancy in the maximum velocity broadening is likely the difference in line strength, leading to more pronounced wings in the strongest lines.  Beyond 12$\,\kms$ the CO line flux falls below the rms noise value of 3 mJy/beam of the observation. In addition, morphological causes may also contribute to the observed difference. Compact differentially rotating disks and wind--companion interactions can provide the required accelerations to generate enhanced line wings \citep{Kervella2016,Homan2017}.

The face-on orientation of the the disk/spiral of EP Aqr suggests that the origin of the enhanced wings is not due to morphology nor that thermal broadening is responsible for the widening. Local CSE temperatures remain below the stellar surface temperature of $\sim$3200 K, which can maximally broaden the line wings of both molecules by 1.5$\,\kms$. We therefore propose either of the two following scenarios:

\begin{itemize}
 \item The complex dynamics in the wind--companion interaction zone not only accelerate a large portion of the outflow material along the equatorial plane, but also along the polar axis, causing some of this material to have increased velocities along the line of sight; 
 \item The broadening of the line wings finds its origin in microscopic interactions, and their width is intrinsically determined by the nature of the stellar outflow. In other words, a measurable, and thus significant portion of the AGB material is in fact accelerated beyond previously measured estimates.
\end{itemize}

These unexpectedly high velocities may be explained by enforced dynamics in an isotropic wind potentially due to the formation of fractal grains up to large radii, as extensively discussed in \citet{Decin2018} for R Dor and IK Tau. We supplement these ideas with the following discussion. The wind acceleration process in single AGB stars is intrinsically chaotic and stochastic, with an extremely complex chain of interactions that transform the original photon momentum distribution into the final gas velocity distribution. This process does not ensure that the original TE nature of the photon momentum distribution remains conserved. Hence, the resulting final radial gas speed is probably a fully non-local thermodynamic equilibrium (NLTE) distribution of velocities whose statistical properties depend on the specific composition of the wind material. Currently, this issue is regularly circumvented by assuming the existence of a globally present Gaussian turbulent velocity field. However, this Gaussian nature reveals the implicit LTE assumption (mimicking the Gaussian nature of a thermal distribution of velocities), which likely does not  hold under the extreme conditions present in AGB winds. This is illustrated by the fact that the broad wings of the observed SiO transition could only be explained by a Gaussian microturbulent velocity distribution with a FWHM of $\sim$4--5$\,\kms$, which is unphysically high.

The study of mixed granular particle fluids has shown that systems subject to inelastic collisions that undergo permanent energy injection can substantially overpopulate the high-velocity tails of the overall particle velocity distribution \citep{VanNoije1998,Soto1999,Pagonabarraga2001,Eshuis2007,Gradenigo2011,Puglisi2012,Pontuale2016}. For mixed fluids the length of this overpopulated tail can be dependent on the particle species \citep{Martin1999,Feitosa2002,Barrat2002,Marconi2002,Pagnani2002}. If AGB winds can be considered   mixed, partly granular fluids harbouring a non-negligible amount of inelastic interactions, the above-mentioned results could explain the observed high-velocity tails. We may argue that the hydrogen bath that composes the bulk of the AGB wind would quickly thermalise all the present gaseous and solid-state components through frequent elastic collisions. However, the occurrence of enough inelastic interactions (in the form of  inelastic collisions with solid-state particles, chemical reactions, or photon absorption) can sustain the NLTE properties of some components in a mixed fluid, even when all are in contact with an equilibrium bath \citep{Feitosa2004,Puglisi2005,Evans2011,Gradenigo2012}.

\section{Summary} \label{summ}

In this paper we have present new ALMA band 6 observations of the circumstellar environment of the AGB star EP Aquarii in cycle 4. The observations were carried out in three different antenna configurations, and probe the molecular emission of the $^{12}$CO $\varv=0$ $J=2-1$, $^{28}$SiO $\varv=0$ $J=5-4$, and SO$_{\rm 2}$ $\varv=0$ $\rm{28}_{\rm 3,25}-\rm{28}_{\rm 2,26}$ transitions. We have discussed the morphological complexities identified in the data.

The CO emission shows an unusual line profile, consisting of a broad plateau with a bright and narrow central peak. The channel maps reveal that the plateau corresponds with the hemispheres of a mostly unperturbed wind, and the peak with a bright nearly face-on spiral structure, making this the first convincing detection of a spiral in the circumstellar environment of an oxygen-rich AGB star. Based on the offsets of the centres of the two unperturbed wind hemispheres, we deduce the position angle of the inclination axis to be $\sim$150$^\circ$ measured anticlockwise from north. Compared to the observed spiral structure in C-rich environments, this spiral is quite different in that it  possesses a high degree of smaller scale hydrodynamical structure and comparatively wide spiral arms. The spiral is extremely confined in velocity-space. We hypothesise that the spiral may either be a radially outflowing density enhancement or a hydrodynamical perturbation in a larger nearly face-on differentially rotating disk. This permits us to estimate the inclination angle of the system to be between 4$^\circ$ and 18$^\circ$; the lowest inclination angles are derived for a radial wind, and the larger ones for a differentially rotating face-on disk. The innermost regions of the CO emission exhibit a morphology that strongly resembles the wind Roche-lobe overflow (WRLOF) simulations of the Mira AB system.

The SiO emission is much smoother than the CO emission, and it does not reveal the spiral. However, it does exhibit a localised emission void, located about 0.5'' west of the continuum brightness peak in the red-shifted portion of the emission. We hypothesise that this feature may be a local environment caused by the presence of a stellar companion, and estimate the mass of the companion to be at most $\sim$0.1 $\mso$ based on its non-detection in the continuum. This tentative companion would be located where the WRLOF simulations predict it to be. A physical connection between the AGB star location and the location of this local environment is detected in CO. 

Finally, the SO$_{\rm 2}$ emission, which remains confined to a radius of $\sim$0.5'', shows a clear sign of (co)rotation, with a maximum projected velocity amplitude of $\sim$1$\,\kms$. Deprojecting these velocities according to the previously deduced inclination angles, the maximum absolute velocity of the gas ranges between 3.5$\,\kms$ (for the disk scenario) and 14$\,\kms$ (for the radial outflow scenario). The odd nature of the spiral and the deprojected rotation speeds of the gas within radii below the companion's orbit favours the hypothesis that the equatorial matter is contained within a large face-on differentially rotating disk, and that the spiral may be an instability in this disk.

The considerable sensitivity of ALMA allows the detection of spectral line wings for SiO whose velocity range is much broader than previously derived from observations. We tentatively attribute these enhanced high-velocity line wings to the potential NLTE nature of the AGB wind, which could be considered a driven, mixed, partly granular fluid subject to occasional inelastic internal interactions.

% ***********************************************************************************************************************************************
%                                                                 END MAIN BODY
% ***********************************************************************************************************************************************

\begin{acknowledgements}  
We would like to thank Dr C. Maes (KU Leuven) for his insightful comments on the NLTE character of certain material mixtures. We would also like to thank Dr S. Mohamed (South African Astronomical Observatory) for her fruitful contributions on the WRLOF mechanism. W.H. acknowledges support from the Fonds voor Wetenschappelijk Onderzoek Vlaanderen (FWO) and support from the ERC consolidator grant 646758 AEROSOL. L.D. acknowledges support from the ERC consolidator grant 646758 AEROSOL. This paper makes use of the following ALMA data: ADS/JAO.ALMA\#2011.0.00277.S. ALMA is a partnership of ESO (representing its member states), NSF (USA), and NINS (Japan), together with NRC (Canada),  NSC and ASIAA (Taiwan), and KASI (Republic of Korea), in cooperation with the Republic of Chile. The Joint ALMA Observatory is operated by ESO, AUI/NR.A.O, and NAOJ.
\end{acknowledgements}

\bibliographystyle{aa}
\bibliography{wardhoman_biblio}

\IfFileExists{wardhoman_biblio.bbl}{}
 {\typeout{}
  \typeout{******************************************}
  \typeout{** Please run "bibtex \jobname" to obtain}
  \typeout{** the bibliography and then re-run LaTeX}
  \typeout{** twice to fix the references!}
  \typeout{******************************************}
  \typeout{}
 }

\cleardoublepage
\begin{appendix}

\section{QSO specifics}

\begin{table*}[]
\centering
\caption{ QSO flux measurements at 224.8 GHz, and their roles. Bp.C. stands for bandpass correction source, P.R. stands for phase reference source. Target J2148+0657 was used as flux standard. \label{qso}}
\begin{tabular}{ cllrr }

\hline
\hline
& QSO & Coordinates & Flux (Jy) & Role  \\
\hline
\multirow{3}{*}{TM2} & J2148+0657 & R.A. 21:48:05.459 Dec. +06:57:38.604 & 0.71 & Bp.C. \\
& J2134-0153 & R.A. 21:34:10.310 Dec. $-$01:53:17.238 & 0.60 & P.R. \\
& J2156-0037 & R.A. 21:56:14.758 Dec. $-$00:37:04.594 & 0.34 & P.R. \\
\hline
\multirow{3}{*}{TM1} & J2148+0657 & R.A. 21:48:05.459 Dec. +06:57:38.604 & 0.69 & Bp.C. \\
& J2134-0153 & R.A. 21:34:10.310 Dec. $-$01:53:17.238 & 0.63 & P.R. \\
& J2156-0037 & R.A. 21:56:14.758 Dec. $-$00:37:04.594 & 0.25 & P.R. \\
\hline
\multirow{2}{*}{ACA} & J2232+1143 & R.A. 22:32:36.409 Dec. +11:43:50.904 & 4.84 & Bp.C. \\
& J2134-0153 & R.A. 21:34:10.310 Dec. $-$01:53:17.238 & 0.67 & P.R. \\

\hline

\end{tabular}
\end{table*}

% \section{Channel maps without contours}
% 
% \begin{figure*}[]
%         \centering
%         \includegraphics[width=17cm]{CO_chanC-eps-converted-to.pdf}
%         \caption{Continuum subtracted channel maps of the CO emission, identical to the channels shown in Fig. \ref{COchancont}. The contours have been omitted to mitigate visual bias.
%         \label{COchan}} 
% \end{figure*}
% 
% \begin{figure*}[]
%         \centering
%         \includegraphics[width=17cm]{SiO_chanC-eps-converted-to.pdf}
%         \caption{Continuum subtracted channel maps of the SiO emission, identical to the channels shown in Fig. \ref{SiOchancont}. The contours have been omitted to mitigate visual bias.
%         \label{SiOchan}} 
% \end{figure*}
% 
% \begin{figure*}[]
%         \centering
%         \includegraphics[width=17cm]{SO2_chanC-eps-converted-to.pdf}
%         \caption{Continuum subtracted channel maps of the SO$_{\rm 2}$ emission, identical to the channels shown in Fig. \ref{SO2chancont}. The contours have been omitted to mitigate visual bias.
%         \label{SO2chan}} 
% \end{figure*}

\section{Zoomed in channel maps of the CO emission}

\begin{figure*}[]
        \centering
        \includegraphics[width=17cm]{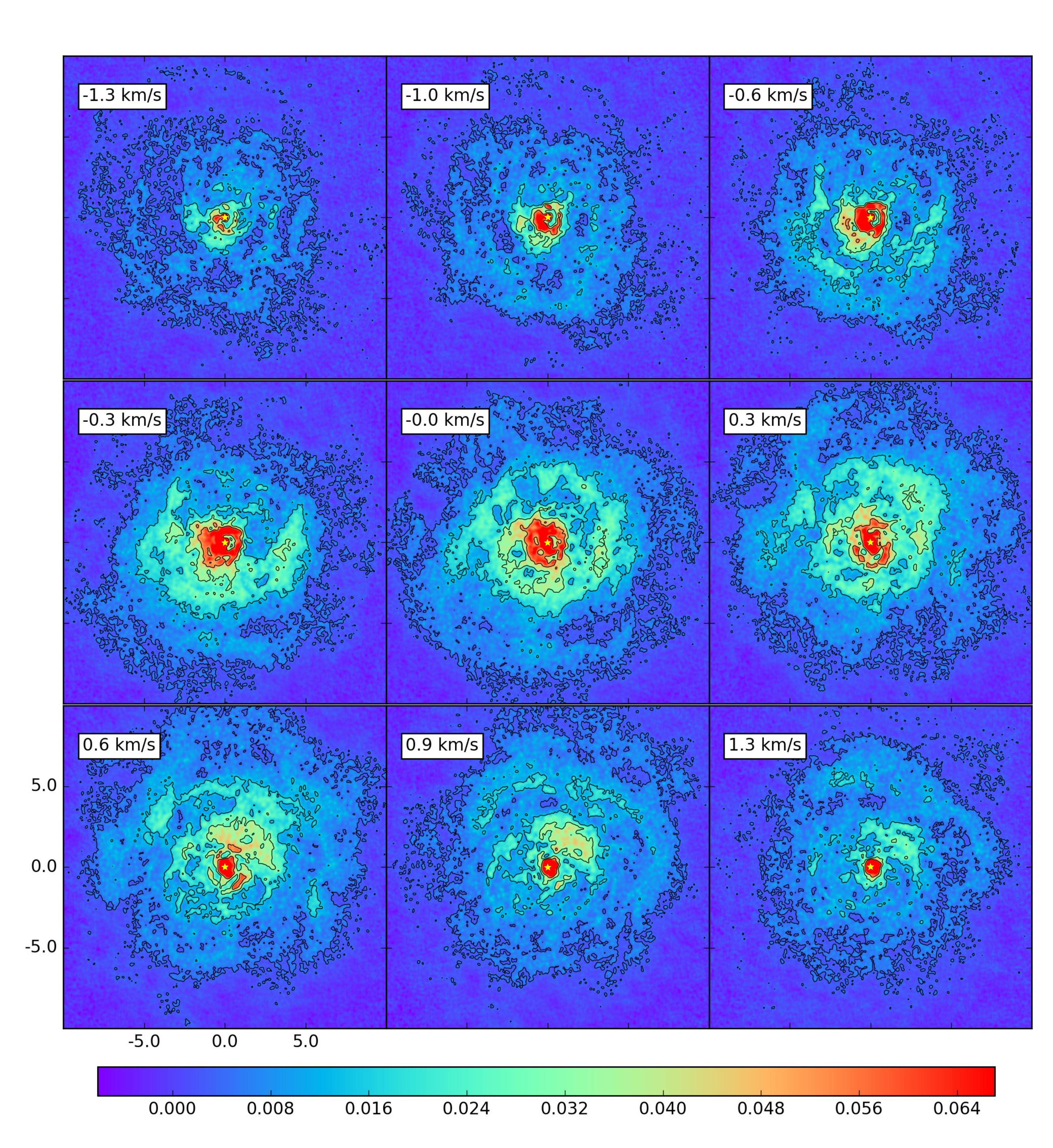}
        \caption{Channel maps of the central peak of the CO spectral line. The contours are drawn at 3, 12, 25, and 40 times the rms noise value outside the line (1.25$\times {\rm 10}^{\rm -3}$ Jy/beam). The continuum brightness peak is indicated with the yellow star. The spiral feature can be seen to appear in the north-west and recede in the south-east, from blue- to red-shifted emission, respectively.
        \label{COzpcont}} 
\end{figure*}

\begin{figure*}[]
        \centering
        \includegraphics[width=17cm]{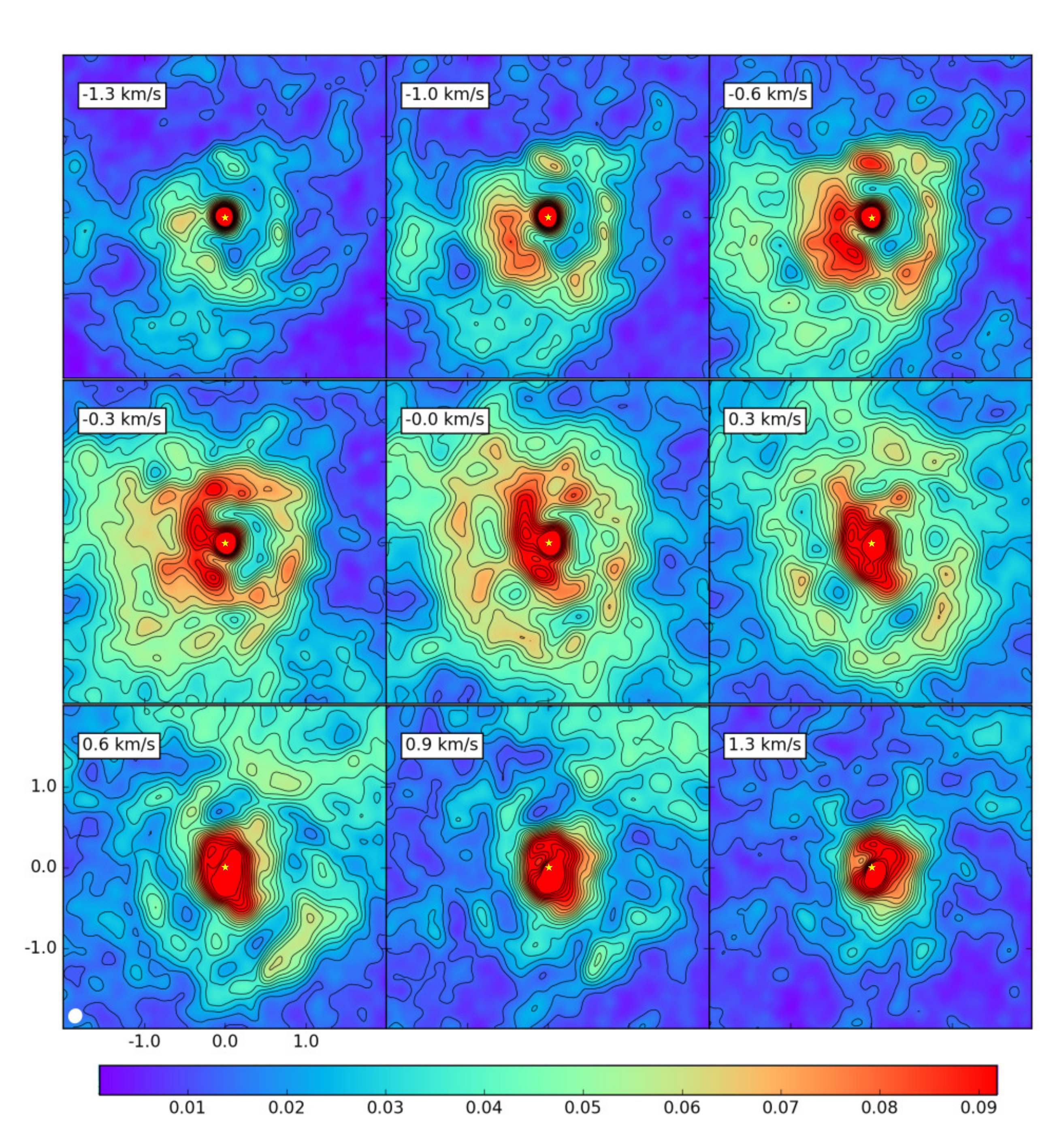}
        \caption{Channel maps of the central peak of the CO spectral line, zooming in on the bright central region. The contours are drawn at 10, 15, 20, ... 85, and 90 times the rms noise value outside the line (1.25$\times {\rm 10}^{\rm -3}$ Jy/beam). The continuum brightness peak is indicated with the yellow star. The density of contours exhibits the high degree of structure in the spiral formation zone.
        \label{COzccont}} 
\end{figure*}

\end{appendix}
 
\end{document}